\documentclass[twocolumn,american,aps, prl, floatfix, table, dvipsnames]{revtex4-2}
\usepackage[T1]{fontenc}
\usepackage[latin9]{inputenc}
\usepackage{color}
\usepackage{babel}
\usepackage{verbatim}
\usepackage{amsmath}
\usepackage{amssymb}
\usepackage{graphicx}
\usepackage{esint}
\usepackage[unicode=true,
 bookmarks=false,
 breaklinks=true,pdfborder={0 0 1},backref=false,colorlinks=true]
 {hyperref}
\hypersetup{
 pdfcreator={},pdfproducer={LaTeX with hyperref},linkcolor=blue,anchorcolor=blue,citecolor=blue,filecolor=red,menucolor=red,pagecolor=red,urlcolor=blue,pdfstartview=FitV,pdfhighlight=/I,pdfpagelayout=TwoColumns,hypertexnames=true}

\makeatletter
\usepackage{lmodern}
\usepackage{mathptmx, newtxtext, newtxmath, xspace}
\usepackage{bbm}
\usepackage{graphicx}

\usepackage{floatrow}

\usepackage{accents}
\newlength{\dhatheight}
\newcommand{\doublehat}[1]{%
    \settoheight{\dhatheight}{\ensuremath{\hat{#1}}}%
    \addtolength{\dhatheight}{-0.15ex}%
    \hat{\vphantom{\rule{1pt}{\dhatheight}}%
    \smash{\hat{#1}}}}

\makeatother

\begin{document}
\title{Local condensation of charge-$4e$ superconductivity at a nematic
domain wall}
\author{Matthias Hecker }
\affiliation{School of Physics and Astronomy, University of Minnesota, Minneapolis
55455 MN, USA}
\author{Rafael M. Fernandes}
\affiliation{School of Physics and Astronomy, University of Minnesota, Minneapolis
55455 MN, USA}
\date{\today }
\begin{abstract}
In the fluctuation regime that precedes the onset of pairing in multi-component
superconductors, such as nematic and chiral superconductors, the normal
state is generally unstable towards the formation of charge-$4e$
order \textendash{} an exotic quantum state in which electrons form
coherent quartets rather than Cooper pairs. However, charge-$4e$
order is often suppressed by other competing composite orders, such
as nematics. Importantly, the formation of nematic domains is unavoidable
due to the long-range strains generated, leading to one-dimensional
regions where the competing nematic order is suppressed. Here, we
employ a real-space variational approach to demonstrate that, in such
nematic domain walls, charge-$4e$ order is locally condensed via
a vestigial-order mechanism. We explore the experimental manifestations
of this effect and discuss materials in which it can be potentially
observed.
\end{abstract}
\maketitle

\paragraph{Introduction.}

Shortly after the development of the BCS model for superconductivity,
it was recognized that a gas of bosons could also form a coherent
state of pairs of bosons before the Bose-Einstein condensation of
individual particles, provided that strong enough attractive interactions
are present \citep{Valatin1958,Evans1969}. In condensed matter systems,
pair condensation of bosonic quasiparticles has been studied in various
settings, from biexcitons in semiconductors \citep{Lampert1958,Moskalenko1958}
to two-magnon bound states in frustrated magnets \citep{Wang2017,Strockoz2022}.
A fascinating possibility is the emergence, in superconducting materials,
of a coherent state of pairs of Cooper pairs, dubbed quartets \citep{Korshunov1985,Volovik1992,Aligia2005,Berg2009,Moon2012,Khalaf2022}.
Theoretically, a charge-$4e$ superconducting state is expected to
display gapless excitations \citep{HongYao2017,Gnezdilov2021} and
half flux-quantum vortices \citep{Berg2009}. Experimentally, charge-$4e$
bound states have been recently invoked to explain puzzling magneto-transport
data in kagome superconductors \citep{Ge2022,Varma2023}.

In the case of bosonic particles, the state with paired bosons is
thermodynamically stable only when there is more than one bosonic
``flavor'' available for condensation (e.g. spin-1 bosons) \citep{Nozieres1982}.
This suggests that ``multi-flavor'' superconductors are a promising
setting to search for charge-$4e$ superconductivity. Indeed, theoretical
proposals for quartet formation have included spin-3/2 systems \citep{Wu2005},
spinor condensates \citep{Moore2006}, multi-band superconductors
\citep{Babaev2010}, pair-density waves \citep{Berg2009,Radzihovsky2009,Radzihovsky2011,Agterberg2011,Agterberg2020,Shaffer2023,Yue2023,Wu_Wang2023},
and multi-component superconductors \citep{Fernandes2021,Jian2021,Zeng2021,Rampp2022,Curtis2023,Scheurer2023}.
In the latter case, the superconductor is described by multiple gap
functions related by lattice symmetries, $\boldsymbol{\Delta}=\left(\Delta_{1},\,\Delta_{2},\cdots\right)$;
in group-theory jargon, $\boldsymbol{\Delta}$ transforms as a multi-dimensional
irreducible representation (IR) of the point group. A broad range
of pairing states belong to this category, including several versions
of $p$-wave and $d$-wave states in tetragonal, hexagonal, and cubic
lattices \citep{Annett1990,Sigrist1991}. More importantly, there
is experimental evidence for the realization of multi-component pairing
in various systems of interest, from heavy-fermions \citep{Sauls1994,Schemm2014,Avers2020}
to moiré superlattices \citep{Cao2021} to doped topological insulators
\citep{Matano2015,Pan2016,Asaba2016}.

Charge-$4e$ order emerges in multi-component superconductors via
the condensation of a \emph{complex-valued }composite order parameter
$\left\langle \boldsymbol{\Delta}\cdot\boldsymbol{\Delta}\right\rangle \neq0$
(distinct from the \emph{real-valued }composite $\left\langle \boldsymbol{\Delta}^{\dagger}\cdot\boldsymbol{\Delta}\right\rangle $),
while the superconducting order parameter itself remains zero, $\left\langle \boldsymbol{\Delta}\right\rangle =0$,
such that the transition temperature of the composite order, $T_{4e}$,
is larger than the superconducting one, $T_{c}$ \citep{Fernandes2021,Hecker2023}.
This spontaneous symmetry-breaking, which is driven by fluctuations
and thus not captured by mean-field approaches, lowers the $U(1)$
gauge symmetry to $Z_{2}$; the latter is further broken if $\left\langle \boldsymbol{\Delta}\right\rangle $
becomes non-zero. It is said then that the charge-$4e$ and charge-$2e$
superconducting states are intertwined, and that the former is a vestigial
order of the latter \citep{Fradkin2015,Fernandes2019}. 

The main obstacle for the stabilization of charge-$4e$ vestigial
order is the competition with other vestigial phases, most notably
nematic and ferromagnetic. Indeed, besides $U(1)$ symmetry, the ground
state of a multi-component superconductor also breaks either time-reversal
(chiral superconductor) or rotational symmetry (nematic superconductor)
\citep{Gali2022}. These additional symmetries can be broken before
the onset of superconductivity via the condensation of \emph{real-valued
}composite order parameters. Large-$N$ and variational calculations
found that the corresponding vestigial nematic and ferromagnetic orders
generally preempt the onset of charge-$4e$ order \citep{Hecker2018,Willa2020}
\textendash{} except for the special case of a hexagonal nematic superconductor
\citep{Fernandes2021}. 

\begin{figure}[t] 
\raggedright{}
\ffigbox[][]{\includegraphics[scale=0.65]{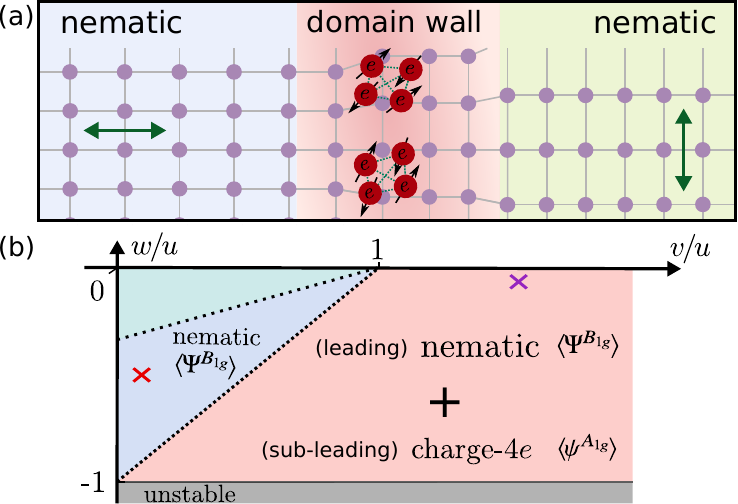}}
{\caption{(a) Real-space sketch of our main result,  the emergence of charge-$4e$ order at a nematic domain wall.  (b) Phase diagram of the vestigial orders supported by the nematic superconducting ground state of a tetragonal system.  Here,  $u$, $v$ and $w$ are the Landau parameters of Eq. (\ref{eq:S_int}).  The different shaded regions indicate which vestigial channel is attractive: none (green),  nematic (blue),  and leading nematic with subleading charge-$4e$ (red). } \label{fig:intro}}      
\end{figure}

In this work, we investigate whether the local suppression of the
leading nematic or ferromagnetic vestigial order enables the local
condensation of charge-$4e$ order. As a motivation, we first discuss
a 2D triplet superconductor without spin-orbit coupling (SOC), whose
multi-component gap function has a continuous SU(2) symmetry. In this
case, the continuous symmetry globally forbids quasi-long-range order
of any composite order parameter, except for charge-$4e$. We then
study the more realistic situation of discrete multi-component superconductors
in tetragonal and hexagonal systems. While long-range order in the
competing vestigial channel is unavoidable, it is locally suppressed
due to the formation of domains. Focusing on a nematic superconductor
on the tetragonal lattice, we employ a real-space variational approach
that treats all composite orders on an equal footing. We find a wide
range of parameters for which charge-$4e$ order is condensed at the
nematic domain walls while the superconducting order parameter remains
uncondensed, as illustrated schematically in Fig. \ref{fig:intro}(a).
We finish by discussing the experimental manifestations of this local
condensation and candidate materials. 

\paragraph{Global suppression of the competing vestigial phases.}

To set the stage, we discuss a special case in which, for symmetry
reasons alone, the only order that can be stabilized is charge-$4e$.
Consider a 2D system in which the electrons experience negligible
SOC (e.g. graphene) and, in addition to spin, have another pseudospin
degree of freedom (e.g. sublattice or valley). Enforcing the pairing
state to be momentum-independent and pseudospin-singlet, the gap function
must be spin-triplet and represented by an order parameter $\boldsymbol{\Delta}=(\Delta_{1},\,\Delta_{2},\,\Delta_{3})^{T}$
that transforms as a vector in SU(2) spin space. This is nothing but
the $\boldsymbol{d}$-vector of a triplet superconductor \citep{Annett1990,Sigrist1991},
albeit even in momentum. While there is no indication of valley-singlet
spin-triplet superconductivity in graphene, this type of state has
been studied in twisted moiré systems \citep{Venderbos_Fernandes,Scheurer2020,Wang_Kang2021,Lake2022}
and Bernal bilayer graphene \citep{Curtis2023,Levitov_Chubukov}.
The superconducting action is given by 
\begin{align}
\mathcal{S} & =\int_{\mathsf{q}}\chi_{\mathsf{q}}^{-1}\left|\boldsymbol{\Delta}_{\mathsf{q}}\right|^{2}+\int_{\mathsf{x}}\Big(u\left|\boldsymbol{\Delta}_{\mathsf{x}}\right|^{4}+v\left|\boldsymbol{\Delta}_{\mathsf{x}}\cdot\boldsymbol{\Delta}_{\mathsf{x}}\right|^{2}\Big),\label{eq:S_triplet}
\end{align}
with (bare) superconducting susceptibility $\chi_{\mathsf{q}}^{-1}$
and $\mathsf{q}=\left(\mathbf{q},\omega_{n}\right)$ denoting momentum
and Matsubara frequency. The interaction part has Landau coefficients
$u$ and $v$, where $\mathsf{x}=\left(\mathbf{x},\tau\right)$ denotes
position and imaginary time. The mean-field phase diagram of this
model is well-established \citep{Volovik1985,Uzunov1990}, displaying
different types of unitary and non-unitary pairing depending on the
sign of $v$.

To describe the vestigial orders, however, it is necessary to go beyond
mean-field \citep{Fernandes2019}; for our purposes, a group-theoretical
analysis is sufficient. There are four symmetry-breaking bilinear
combinations of $\boldsymbol{\Delta}$: two real-valued composites
$\boldsymbol{\Psi}^{(l=1)}=\mathsf{i}\boldsymbol{\Delta}\times\bar{\boldsymbol{\Delta}}$
and $\Psi_{\mu\nu}^{(l=2)}=\frac{1}{2}\left(\Delta_{\mu}\bar{\Delta}_{\nu}+\Delta_{\nu}\bar{\Delta}_{\mu}\right)-\frac{1}{3}\,\delta_{\mu\nu}\left|\boldsymbol{\Delta}\right|^{2}$
and two complex-valued ones, $\psi^{(l=0)}=\boldsymbol{\Delta}\cdot\boldsymbol{\Delta}$
and $\psi_{\mu\nu}^{(l=2)}=\Delta_{\mu}\Delta_{\nu}-\frac{1}{3}\,\delta_{\mu\nu}\left(\boldsymbol{\Delta}\cdot\boldsymbol{\Delta}\right)$,
where $\bar{z}\equiv z^{*}$. The superscript indicates the transformation
properties in SU(2) spin-space, corresponding to a scalar ($l=0$),
a vector ($l=1$), or a tensor ($l=2$). Each composite has a clear
physical interpretation: $\boldsymbol{\Psi}^{(l=1)}$ corresponds
to time-reversal symmetry-breaking and, thus, vestigial ferromagnetic
order. $\Psi_{\mu\nu}^{(l=2)}$ is associated with rotational symmetry-breaking
in spin-space, and therefore denotes (spin-)nematic vestigial order
\citep{Blume1969,Andreev1984,Rampp2022}. Finally, $\psi^{(l=0)}$
and $\psi_{\mu\nu}^{(l=2)}$ correspond to $s$-wave and $d$-wave
charge-$4e$ vestigial orders, respectively.

Because $\boldsymbol{\Psi}^{(l=1)}$, $\Psi_{\mu\nu}^{(l=2)}$, $\psi_{\mu\nu}^{(l=2)}$,
and $\boldsymbol{\Delta}\equiv\boldsymbol{\Delta}^{(l=1)}$ itself
transform non-trivially in SU(2) spin-space (i.e. they are at least
Heisenberg-type order parameters), none of them can sustain (quasi-)long-range
order at non-zero temperatures in 2D. On the other hand, because $\psi^{(l=0)}$
is a complex scalar (i.e. XY-type), it can establish quasi-long-range
order through a BKT transition. Therefore, the only state allowed
to develop quasi-long-range order in this model is the vestigial charge-$4e$
phase. We note that similar 2D and 1D models for triplet superconductors
\citep{Rampp2022,Curtis2023,Scheurer2023} and spinor condensates
\citep{Moore2006} have been previously studied and shown to support
charge-$4e$ order.

\paragraph{Local suppression of the competing vestigial phases.}

Despite illuminating, the simple model above is not representative
of realistic multi-component superconductors, where either SOC is
not negligible or singlet states are realized. Yet, it highlights
an efficient strategy to realize charge-$4e$ order: suppression of
the other, leading, vestigial phases. While this generally cannot
be accomplished globally via Mermin-Wagner's theorem, vestigial nematic
or ferromagnetic states tend to form domains to minimize the elastic
or magnetic dipolar energies. To explore this idea, we consider a
generic two-component superconducting order parameter $\boldsymbol{\Delta}=(\Delta_{1},\Delta_{2})$,
which could describe $(p_{x},p_{y})$-wave or $(d_{xz},d_{yz})$-wave
states in tetragonal and hexagonal lattices \citep{Annett1990,Sigrist1991},
or $(d_{x^{2}-y^{2}},d_{xy})$-wave pairing in hexagonal systems and
$45^{\circ}$-twisted bilayer tetragonal $d$-wave superconductors
\citep{Franz2022}. In contrast to the previous example, $\boldsymbol{\Delta}$
now transforms as a two-dimensional IR of a discrete point group.
While we will focus on tetragonal ($\mathrm{D_{4h}}$) superconductors,
where $\boldsymbol{\Delta}$ transforms as the IR $E_{g}$ or $E_{u}$,
the conclusions apply to all other cases.

We start by classifying all possible composite order parameters. As
shown in \citep{Hecker2023}, seven different bilinear combinations
can be formed. Apart from the symmetry-preserving bilinear $\Psi^{A_{1g}}=\boldsymbol{\Delta}^{\dagger}\tau^{0}\boldsymbol{\Delta}$,
with Pauli matrices $\tau^{i}$ acting on the $\boldsymbol{\Delta}$
subspace, there are three additional real-valued bilinears: 
\begin{align}
\Psi^{A_{2g}} & =\boldsymbol{\Delta}^{\dagger}\tau^{y}\boldsymbol{\Delta}, & \Psi^{B_{1g}} & =\boldsymbol{\Delta}^{\dagger}\tau^{z}\boldsymbol{\Delta}, & \Psi^{B_{2g}} & =\boldsymbol{\Delta}^{\dagger}\tau^{x}\boldsymbol{\Delta}.\label{eq:Psis}
\end{align}
Here, $\Psi^{A_{2g}}$ breaks time-reversal-symmetry and causes ferromagnetism,
while $\Psi^{B_{1g}}$ and $\Psi^{B_{2g}}$ break tetragonal symmetry
and cause nematicity. The three complex-valued bilinears are given
by 
\begin{align}
\psi^{A_{1g}} & =\boldsymbol{\Delta}^{T}\tau^{0}\boldsymbol{\Delta}, & \psi^{B_{1g}} & =\boldsymbol{\Delta}^{T}\tau^{z}\boldsymbol{\Delta}, & \psi^{B_{2g}} & =\boldsymbol{\Delta}^{T}\tau^{x}\boldsymbol{\Delta},\label{eq:psis}
\end{align}
and describe, respectively, $s$-wave, $d_{x^{2}-y^{2}}$-wave, and
$d_{xy}$-wave charge-$4e$ superconductivity. In our notation, $\Psi^{n}$
($\psi^{n}$) denotes real-valued (complex-valued) bilinears, whereas
the superscript $n$ indicates the IR according to which the composite
transforms. The superconducting action is $\mathcal{S}=\int_{\mathsf{x}}r_{0}\left|\boldsymbol{\Delta}_{\mathsf{x}}\right|^{2}+\;\mathcal{S}^{\mathrm{grad}}\;+\;\mathcal{S}^{\mathrm{int}}\,,$
where $r_{0}=a_{0}(T-T_{0})$ denotes the bare SC transition ($a_{0},T_{0}>0$)
and $\mathcal{S}^{\mathrm{grad}}$ contains the symmetry-allowed gradient
terms \citep{Sigrist1991}. The interaction part is given by \citep{Hecker2023}
\begin{align}
\mathcal{S}^{\mathrm{int}} & =\int_{\mathsf{x}}\Big[u\,(\Psi_{\mathsf{x}}^{A_{1g}})^{2}+v\,(\Psi_{\mathsf{x}}^{A_{2g}})^{2}+w\,(\Psi_{\mathsf{x}}^{B_{1g}})^{2}\Big],\label{eq:S_int}
\end{align}
and contains three independent interaction parameters $u>0$ and $v,w>-u$.
The mean-field phase diagram in the $\left(\frac{v}{u},\,\frac{w}{u}\right)$
parameter space is well-established, displaying chiral and two types
of nematic superconductivity \citep{Sigrist1991,Berg2009,Gali2022}. 

The vestigial orders associated with each mean-field ground state
were analyzed in Ref. \citep{Hecker2023} via a variational approach.
The leading vestigial instability is always that of a real-valued
composite (nematic or ferromagnetic) whereas the vestigial charge-$4e$
orders are always subleading. While our conclusions hold across the
entire phase diagram, hereafter we focus on the $v>0>w$ region {[}Fig.
\ref{fig:intro}(b){]}, where the superconducting ground state is
nematic and the competing vestigial phases are $d_{x^{2}-y^{2}}$
nematic ($\Psi^{B_{1g}}$) and $s$-wave charge-$4e$ ($\psi^{A_{1g}}$).
In bulk, the only vestigial order realized is the nematic one \citep{Hecker2023}.
However, because the effective interaction in the charge-$4e$ channel
is attractive, the system could gain energy by condensing this mode
in regions where nematic order is suppressed. Due to its Ising-like
character, $\Psi^{B_{1g}}$ can sustain long-range order at non-zero
temperatures. However, because of the linear coupling between $\Psi^{B_{1g}}$
and strain, nematic domains must form to minimize the elastic energy
and accommodate long-range lattice deformations. This opens up the
possibility of $\psi^{A_{1g}}$ condensation at nematic domain walls.

\paragraph{Charge-4e condensation at the domain wall.}

To proceed, we employ a real-space Gaussian variational approach,
which treats all vestigial channels equally, on a 1D grid of length
$L$ and $i=1,\dots,N$ sites. The variational ansatz consists of
a trial action $\mathcal{S}_{0}$ \citep{Moshe2003,Fischer2016,Nie2017,Hecker2023},
which in our case is $\mathcal{S}_{0}=\frac{1}{2}\frac{L}{T}\sum_{i}\hat{\boldsymbol{\Delta}}_{i}^{\dagger}\,G_{i}^{-1}\,\hat{\boldsymbol{\Delta}}_{i}+\mathcal{S}^{\mathrm{grad}}$.
Represented in the Nambu basis $\hat{\boldsymbol{\Delta}}_{i}=(\boldsymbol{\Delta}_{i},\bar{\boldsymbol{\Delta}}_{i})$,
the local inverse Green's function\begin{align} G_{i}^{-1} & =
\left(\begin{smallmatrix} 
R_{i}+\Phi_{i}^{B_{1g}} & \Phi_{i}^{B_{2g}}-\mathsf{i}\Phi_{i}^{A_{2g}} & \phi_{i}^{A_{1g}}+\phi_{i}^{B_{1g}} & \phi_{i}^{B_{2g}}\\  
& R_{i}-\Phi_{i}^{B_{1g}} & \phi_{i}^{B_{2g}} & \phi_{i}^{A_{1g}}-\phi_{i}^{B_{1g}}\\  
&  & R_{i}+\Phi_{i}^{B_{1g}} & \Phi_{i}^{B_{2g}}+\mathsf{i}\Phi_{i}^{A_{2g}}\\ 
\mathrm{H.c.} &  &  & R_{i}-\Phi_{i}^{B_{1g}} 
\end{smallmatrix}\right),
\label{eq:var_Greens_fcn2} 
\end{align}%
contains the real-valued ($\Phi_{i}^{n}$) and complex-valued $(\phi_{i}^{n},\bar{\phi}_{i}^{n})$
variational parameters, and the mass renormalization parameter $R_{i}=r_{0}+\Phi_{i}^{A_{1g}}$.
Because $\mathcal{S}_{0}$ is Gaussian, it is straightforward to compute
the variational free energy
\begin{align}
F_{v} & =-T\log Z_{0}+T\langle\mathcal{S}-\mathcal{S}_{0}\rangle_{0},\label{eq:Fv_0}
\end{align}
where $Z_{0}=\int D(\boldsymbol{\Delta},\bar{\boldsymbol{\Delta}})\,e^{-\mathcal{S}_{0}}$
is the partition function of the trial action. The detailed evaluation
of Eq. (\ref{eq:Fv_0}) is shown in the Supplementary Material (SM).
Importantly, the original Landau coefficients $u$, $v$, and $w$
appear in $F_{v}$ as different combinations in each symmetry channel,
corresponding to effective interactions $U_{A_{1g}}=3u+v+w$, $U_{A_{2g}}=u+3v-w,$
$U_{B_{1g}}=u-v+3w$, $U_{B_{2g}}=u-v-w$, $u_{A_{1g}}=u-v+w$, $u_{B_{1g}}=u+v+w$,
and $u_{B_{2g}}=u+v-w$ \citep{Hecker2023}. An attractive interaction
($U_{n},u_{n}<0$) indicates a potential instability, signaled by
the condensation of the corresponding variational order parameter
($\Phi_{i}^{n},\,\phi_{i}^{n}$). Importantly, a vestigial phase only
emerges if superconductivity is not immediately triggered by $\Phi_{i}^{n},\,\phi_{i}^{n}\neq0$.
In the variational approach, this condition can be verified by confirming
that the variational superconducting susceptibility $\chi$, which
is given by a combination of $R_{i}$, $\Phi_{i}^{n}$, and $\phi_{i}^{n}$
(see SM for details), remains finite. 

\begin{figure}[t]
\includegraphics[scale=0.36]{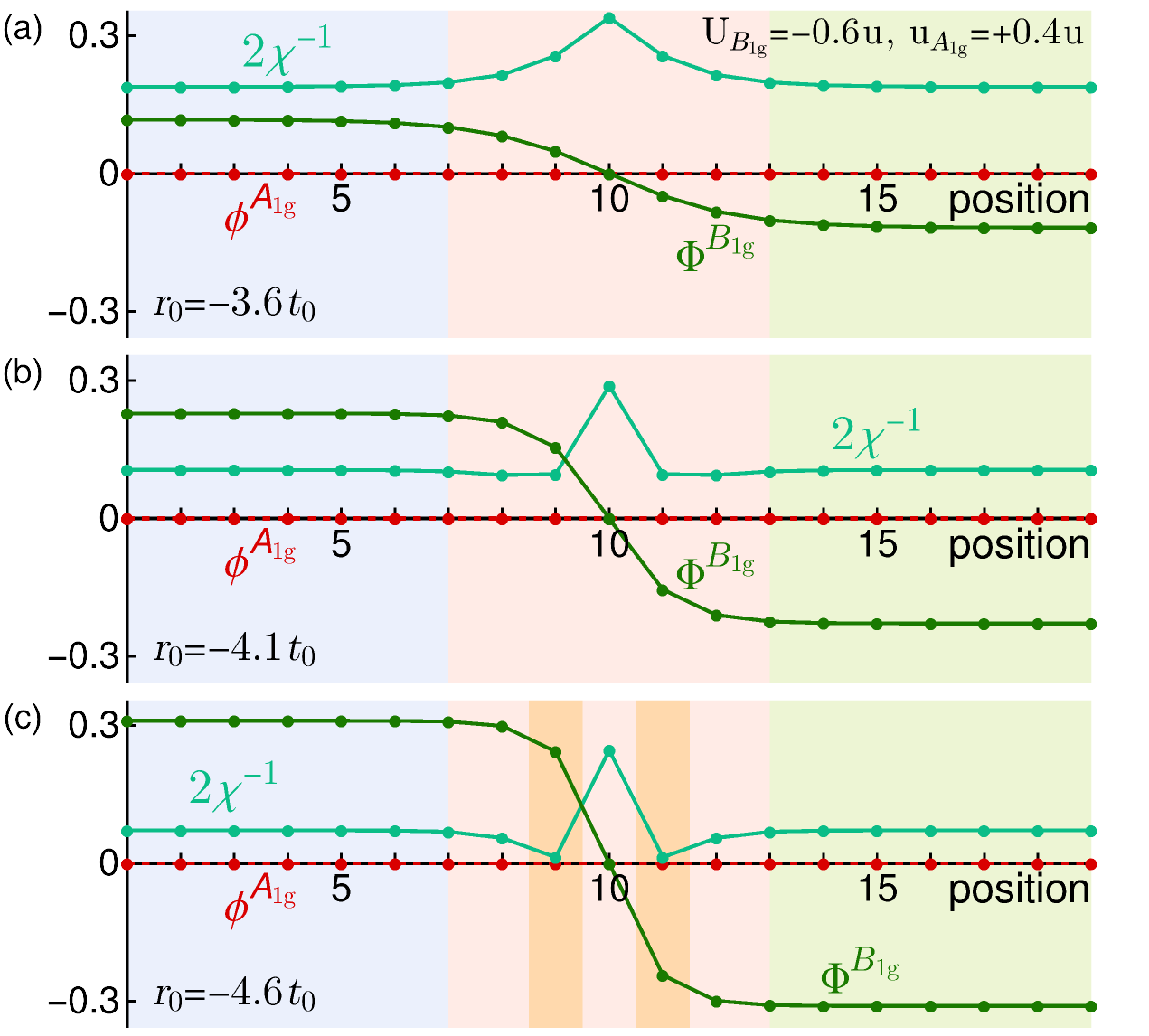}

\caption{Variational nematic order parameter ($\Phi^{B_{1g}}$), inverse superconducting
susceptibility ($\chi^{-1}$), and variational charge-$4e$ order
parameter ($\phi^{A_{1g}}$) obtained from the numerical minimization
of the variational free energy (\ref{eq:Fv_0}) across two nematic
domains (all in units of the gradient-term stiffness $t_{0}$). Each
panel corresponds to a different temperature, parametrized by $r_{0}\propto T-T_{0}$.
The parameters used here are $w=-0.5u$ and $v=0.1u$ {[}red cross
in Fig. \ref{fig:intro}(b){]}. Here and in Fig. \ref{fig:temp_evol},
we set $t_{0}/\sqrt{uT_{0}/L}=20/7$. \label{fig:PureNem}}
\end{figure}

\begin{figure}[t]
\includegraphics[scale=0.36]{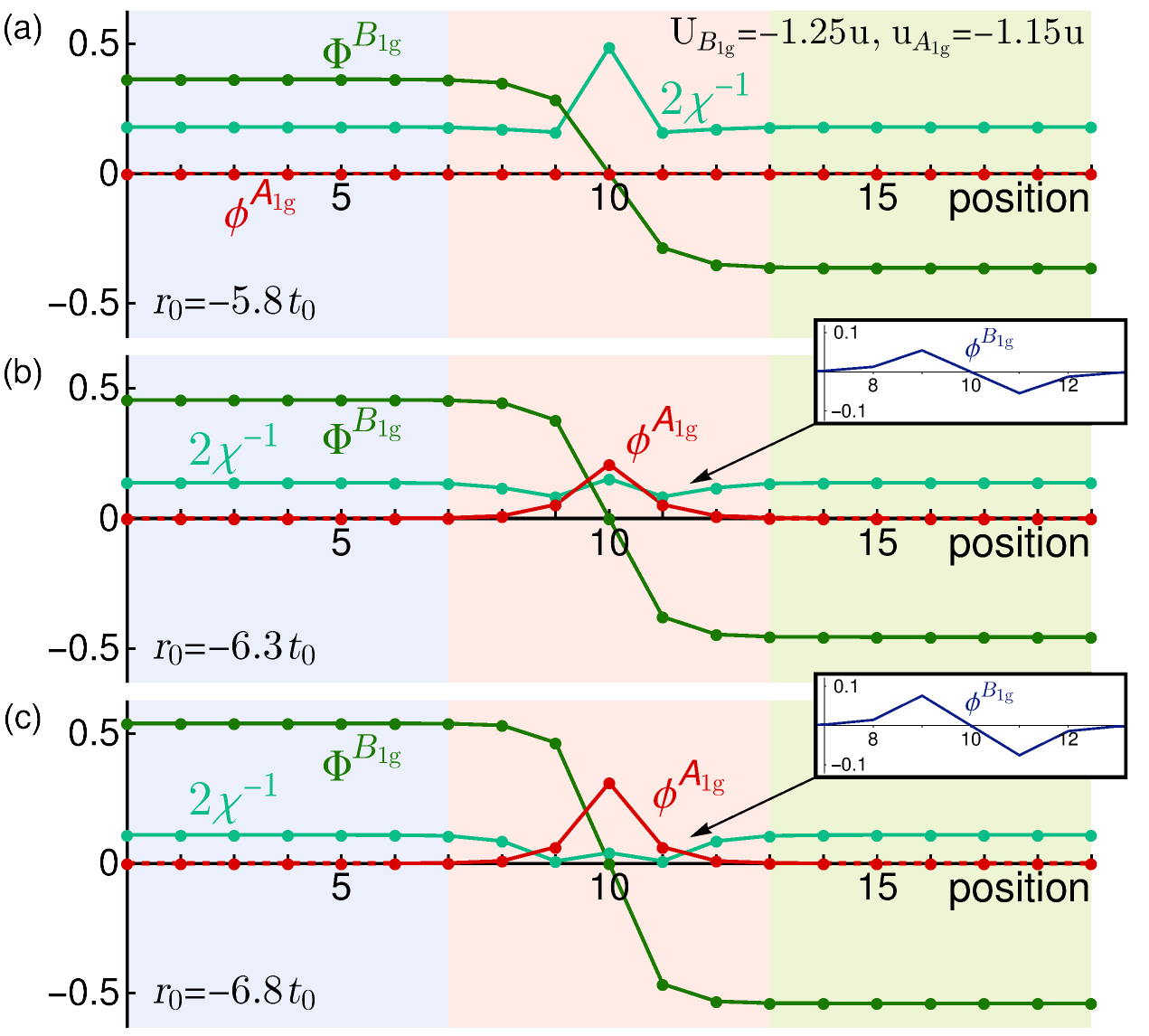}

\caption{Same variational parameters as in Fig. \ref{fig:PureNem}, but evaluated
for the parameters $w=-0.05u$ and $v=2.1u$ {[}purple cross in Fig.
\ref{fig:intro}(b){]}, corresponding to a subleading attractive charge-$4e$
channel ($u_{A_{1g}}<0$). Local condensation of $\phi^{A_{1g}}$
is observed at the nematic domain wall. \label{fig:temp_evol}}
\end{figure}

In bulk, for most of the $\left(\frac{v}{u},\,\frac{w}{u}\right)$
phase diagram, there are two competing attractive vestigial-order
channels, corresponding to a real-valued composite (nematic/ferromagnetic)
and a complex-valued one ($s$-wave/$d$-wave charge-$4e$). As demonstrated
in Ref. \citep{Hecker2023}, the nematic/ferromagnetic instability
is always the leading one in bulk {[}Fig. \ref{fig:intro}(b){]}.
Our goal is to determine the fate of these phases along a nematic
domain wall. We therefore numerically minimize the free energy (\ref{eq:Fv_0})
for the $N$-site 1D grid with domain-wall boundary conditions $\Phi_{1}^{B_{1g}}=-\Phi_{N}^{B_{1g}}=\Phi_{0}^{B_{1g}}$,
where $\Phi_{0}^{B_{1g}}$ is the (self-consistently obtained) bulk
value of the nematic order parameter (see SM for additional details).
Consider first the phase diagram region where the only attractive
vestigial-order channel is the nematic, i.e. $U_{B_{1g}}<0$ but $u_{A_{1g}}>0$
{[}red cross in Fig. \ref{fig:intro}(b){]}. The results are shown
in Fig. \ref{fig:PureNem}. As the control parameter $r_{0}\propto T-T_{0}$
is decreased, a nematic domain emerges below a temperature that coincides
with the bulk nematic critical temperature. As expected, the domain
wall becomes sharper as the temperature is lowered, since the wall
width scales as $t_{0}/|\Phi_{0}^{B_{1g}}|$ where $t_{0}$ is the
gradient-term stiffness. At the domain-wall center, the superconducting
susceptibility $\chi$ is suppressed, consistent with the fact that
vestigial order enhances the superconducting transition. Interestingly,
$\chi^{-1}$ has a non-monotonic spatial dependence, displaying a
dip at the domain-wall boundaries, which can lead to local condensation
of superconductivity {[}yellow region of Fig. \ref{fig:PureNem}(c){]}.
No sign of charge-$4e$ order is observed, consistent with a repulsive
effective interaction in this channel.

Consider now the phase-diagram region where the charge-$4e$ vestigial
channel is attractive, but subleading to the nematic, $U_{B_{1g}}<u_{A_{1g}}<0$
{[}purple cross in Fig. \ref{fig:intro}(b){]} . As shown in Fig.
\ref{fig:temp_evol}, upon decreasing $r_{0}$, nematic order emerges
first, in two domains. However, in contrast to Fig. \ref{fig:PureNem},
the charge-$4e$ order parameter $\phi^{A_{1g}}$ condenses inside
the domain wall while the superconducting susceptibility $\chi$ remains
finite. This is the main result of the paper. Because the domain wall
is one-dimensional, this should be understood as a local condensation,
since phase slips will destroy long-range order along the wall. Note,
even though the $d$-wave charge-$4e$ channel is repulsive in this
phase-diagram region, the order parameter $\phi^{B_{1g}}$ condenses
due to the trilinear coupling between $\Phi^{B_{1g}}$, $\phi^{A_{1g}}$,
and $\bar{\phi}^{B_{1g}}$ (see the inset) \citep{Hecker2023}. We
verified that this behavior is not particular to this set of parameters,
but occurs in any region of the $\left(\frac{v}{u},\,\frac{w}{u}\right)$
phase diagram where $s$-wave/$d$-wave vestigial charge-$4e$ order
is sufficiently attractive but subleading to nematic/ferromagnetic
vestigial order. 

\paragraph{Discussion.}

In this work, we employed a variational approach to demonstrate that
charge-$4e$ order locally condenses at nematic domain walls that
emerge in the normal-state of nematic superconductors upon approaching
the superconducting transition. Before the onset of superconductivity,
the system is unstable towards the formation of vestigial nematic
and charge-$4e$ orders arising from the fluctuation-induced condensation
of composite superconducting order parameters. Because nematic order
generally wins over charge-$4e$ order, the suppression of nematic
order along the domain wall enables the system to further minimize
the free energy by condensing charge-$4e$ order. Since an analogous
mechanism also applies for chiral superconductors, our results point
to a wide class of systems \textendash{} multi-component superconductors
\textendash{} where local charge-$4e$ order can potentially emerge.
Among the materials for which there is strong experimental evidence
for nematic superconductivity, doped $\mathrm{Bi_{2}Se_{3}}$ \citep{Fu2014,Matano2015,Asaba2016,Pan2016}
and twisted bilayer graphene (TBG) \citep{Cao2021,Kozii2019,Chichinadze2020}
are the most promising candidates to display this effect. Indeed,
a vestigial nematic phase exists in doped $\mathrm{Bi_{2}Se_{3}}$
\citep{Tamegai2019,Cho2019}, whereas in TBG, normal-state nematic
order appears close to the superconducting dome \citep{Cao2021}.
There are also several chiral-superconductor candidates \citep{Ghosh2020},
including the widely studied heavy-fermion UPt$_{3}$ \citep{Sauls1994,Schemm2014,Avers2020}.

An important question is how to experimentally detect this effect.
Since charge-$4e$ order emerges at nematic domain walls, local probes
such as scanning tunneling microscopy (STM) and scanning near-field
optical microscopy (SNOM) are ideal. Because the charge-$4e$ state
is expected to be gapless \citep{HongYao2017,Gnezdilov2021}, its
density-of-states (DOS) profile, which can be accessed in a standard
STM measurement, will likely differ from the normal-state DOS only
in subtle ways. On the other hand, Josephson-STM, in which a superconducting
tip is used \citep{Naaman2001,Allan2019}, could provide more direct
evidence for quartets. The SNOM technique has the unique capability
of probing the local optical response with a few nanometers resolution, from which the properties of
the local optical conductivity $\sigma(\omega,\boldsymbol{r})$ can
be inferred \citep{Yang2013,Mcleod2017}. Because the charge-$4e$
state has a non-zero superfluid density \citep{Gnezdilov2021}, $\mathrm{Im}\,\sigma(\omega,\boldsymbol{r})\sim1/\omega$
is expected to emerge at low frequencies near nematic domain walls.
While this behavior could also be due to a charge-$2e$ superconducting
filament, only in the charge-$4e$ case this behavior would be accompanied
by a gapless DOS, which can be probed via STM. These results also
reveal the tantalizing possibility of using uniaxial strain to control
the charge-$4e$ phase, since beyond a critical strain value, the
sample becomes mono-nematic-domain and local charge-$4e$ order disappears. 
\begin{acknowledgments}
We thank J. Schmalian, L. Fu, and R. Willa for fruitful discussions,
and particularly Y. Wang for in-depth discussions about the properties
of charge-$4e$ superconductors. This work was supported by the U.
S. Department of Energy, Office of Science, Basic Energy Sciences,
Materials Sciences and Engineering Division, under Award No. DE-SC0020045.
\end{acknowledgments}

\bibliography{../Literature/LocalCharge4e}

\begin{thebibliography}{71}%
\makeatletter
\providecommand \@ifxundefined [1]{%
 \@ifx{#1\undefined}
}%
\providecommand \@ifnum [1]{%
 \ifnum #1\expandafter \@firstoftwo
 \else \expandafter \@secondoftwo
 \fi
}%
\providecommand \@ifx [1]{%
 \ifx #1\expandafter \@firstoftwo
 \else \expandafter \@secondoftwo
 \fi
}%
\providecommand \natexlab [1]{#1}%
\providecommand \enquote  [1]{``#1''}%
\providecommand \bibnamefont  [1]{#1}%
\providecommand \bibfnamefont [1]{#1}%
\providecommand \citenamefont [1]{#1}%
\providecommand \href@noop [0]{\@secondoftwo}%
\providecommand \href [0]{\begingroup \@sanitize@url \@href}%
\providecommand \@href[1]{\@@startlink{#1}\@@href}%
\providecommand \@@href[1]{\endgroup#1\@@endlink}%
\providecommand \@sanitize@url [0]{\catcode `\\12\catcode `\$12\catcode
  `\&12\catcode `\#12\catcode `\^12\catcode `\_12\catcode `\%12\relax}%
\providecommand \@@startlink[1]{}%
\providecommand \@@endlink[0]{}%
\providecommand \url  [0]{\begingroup\@sanitize@url \@url }%
\providecommand \@url [1]{\endgroup\@href {#1}{\urlprefix }}%
\providecommand \urlprefix  [0]{URL }%
\providecommand \Eprint [0]{\href }%
\providecommand \doibase [0]{https://doi.org/}%
\providecommand \selectlanguage [0]{\@gobble}%
\providecommand \bibinfo  [0]{\@secondoftwo}%
\providecommand \bibfield  [0]{\@secondoftwo}%
\providecommand \translation [1]{[#1]}%
\providecommand \BibitemOpen [0]{}%
\providecommand \bibitemStop [0]{}%
\providecommand \bibitemNoStop [0]{.\EOS\space}%
\providecommand \EOS [0]{\spacefactor3000\relax}%
\providecommand \BibitemShut  [1]{\csname bibitem#1\endcsname}%
\let\auto@bib@innerbib\@empty
\bibitem [{\citenamefont {Valatin}\ and\ \citenamefont
  {Butler}(1958)}]{Valatin1958}%
  \BibitemOpen
  \bibfield  {author} {\bibinfo {author} {\bibfnamefont {J.}~\bibnamefont
  {Valatin}}\ and\ \bibinfo {author} {\bibfnamefont {D.}~\bibnamefont
  {Butler}},\ }\bibfield  {title} {\bibinfo {title} {On the collective
  properties of a boson system},\ }\href@noop {} {\bibfield  {journal}
  {\bibinfo  {journal} {Nuovo Cimento}\ }\textbf {\bibinfo {volume} {10}},\
  \bibinfo {pages} {37} (\bibinfo {year} {1958})}\BibitemShut {NoStop}%
\bibitem [{\citenamefont {Evans}\ and\ \citenamefont {Imry}(1969)}]{Evans1969}%
  \BibitemOpen
  \bibfield  {author} {\bibinfo {author} {\bibfnamefont {W.}~\bibnamefont
  {Evans}}\ and\ \bibinfo {author} {\bibfnamefont {Y.}~\bibnamefont {Imry}},\
  }\bibfield  {title} {\bibinfo {title} {On the pairing theory of the bose
  superfluid},\ }\href@noop {} {\bibfield  {journal} {\bibinfo  {journal}
  {Nuovo Cimento B}\ }\textbf {\bibinfo {volume} {63}},\ \bibinfo {pages} {155}
  (\bibinfo {year} {1969})}\BibitemShut {NoStop}%
\bibitem [{\citenamefont {Lampert}(1958)}]{Lampert1958}%
  \BibitemOpen
  \bibfield  {author} {\bibinfo {author} {\bibfnamefont {M.~A.}\ \bibnamefont
  {Lampert}},\ }\bibfield  {title} {\bibinfo {title} {Mobile and immobile
  effective-mass-particle complexes in nonmetallic solids},\ }\href
  {https://doi.org/10.1103/PhysRevLett.1.450} {\bibfield  {journal} {\bibinfo
  {journal} {Phys. Rev. Lett.}\ }\textbf {\bibinfo {volume} {1}},\ \bibinfo
  {pages} {450} (\bibinfo {year} {1958})}\BibitemShut {NoStop}%
\bibitem [{\citenamefont {Moskalenko}(1958)}]{Moskalenko1958}%
  \BibitemOpen
  \bibfield  {author} {\bibinfo {author} {\bibfnamefont {S.~A.}\ \bibnamefont
  {Moskalenko}},\ }\href@noop {} {\bibfield  {journal} {\bibinfo  {journal}
  {Opt. Spektrosk}\ }\textbf {\bibinfo {volume} {5}},\ \bibinfo {pages} {147}
  (\bibinfo {year} {1958})}\BibitemShut {NoStop}%
\bibitem [{\citenamefont {Wang}\ \emph {et~al.}(2017)\citenamefont {Wang},
  \citenamefont {Feiguin}, \citenamefont {Zhu}, \citenamefont {Starykh},
  \citenamefont {Chubukov},\ and\ \citenamefont {Batista}}]{Wang2017}%
  \BibitemOpen
  \bibfield  {author} {\bibinfo {author} {\bibfnamefont {Z.}~\bibnamefont
  {Wang}}, \bibinfo {author} {\bibfnamefont {A.~E.}\ \bibnamefont {Feiguin}},
  \bibinfo {author} {\bibfnamefont {W.}~\bibnamefont {Zhu}}, \bibinfo {author}
  {\bibfnamefont {O.~A.}\ \bibnamefont {Starykh}}, \bibinfo {author}
  {\bibfnamefont {A.~V.}\ \bibnamefont {Chubukov}},\ and\ \bibinfo {author}
  {\bibfnamefont {C.~D.}\ \bibnamefont {Batista}},\ }\bibfield  {title}
  {\bibinfo {title} {Chiral liquid phase of simple quantum magnets},\ }\href
  {https://doi.org/10.1103/PhysRevB.96.184409} {\bibfield  {journal} {\bibinfo
  {journal} {Phys. Rev. B}\ }\textbf {\bibinfo {volume} {96}},\ \bibinfo
  {pages} {184409} (\bibinfo {year} {2017})}\BibitemShut {NoStop}%
\bibitem [{\citenamefont {Strockoz}\ \emph {et~al.}(2022)\citenamefont
  {Strockoz}, \citenamefont {Antonenko}, \citenamefont {LaBelle},\ and\
  \citenamefont {Venderbos}}]{Strockoz2022}%
  \BibitemOpen
  \bibfield  {author} {\bibinfo {author} {\bibfnamefont {J.}~\bibnamefont
  {Strockoz}}, \bibinfo {author} {\bibfnamefont {D.~S.}\ \bibnamefont
  {Antonenko}}, \bibinfo {author} {\bibfnamefont {D.}~\bibnamefont {LaBelle}},\
  and\ \bibinfo {author} {\bibfnamefont {J.~W.}\ \bibnamefont {Venderbos}},\
  }\bibfield  {title} {\bibinfo {title} {Excitonic instability towards a
  {Potts}-nematic quantum paramagnet},\ }\href@noop {} {\bibfield  {journal}
  {\bibinfo  {journal} {arXiv:2211.11739}\ } (\bibinfo {year}
  {2022})}\BibitemShut {NoStop}%
\bibitem [{\citenamefont {Korshunov}(1985)}]{Korshunov1985}%
  \BibitemOpen
  \bibfield  {author} {\bibinfo {author} {\bibfnamefont {S.}~\bibnamefont
  {Korshunov}},\ }\bibfield  {title} {\bibinfo {title} {Two-dimensional
  superfluid {Fermi} liquid with $p$-pairing},\ }\href
  {http://serkor.itp.ac.ru/pdf//85-2D_Fermi_liquid_with_p-pairing.pdf}
  {\bibfield  {journal} {\bibinfo  {journal} {Zh. Eksp. Teor. Fiz}\ }\textbf
  {\bibinfo {volume} {89}},\ \bibinfo {pages} {539} (\bibinfo {year}
  {1985})}\BibitemShut {NoStop}%
\bibitem [{\citenamefont {Volovik}(1992)}]{Volovik1992}%
  \BibitemOpen
  \bibfield  {author} {\bibinfo {author} {\bibfnamefont {G.~E.}\ \bibnamefont
  {Volovik}},\ }\href {https://doi.org/10.1142/1439} {\emph {\bibinfo {title}
  {Exotic properties of superfluid {3He}}}},\ Vol.~\bibinfo {volume} {1}\
  (\bibinfo  {publisher} {World Scientific},\ \bibinfo {year}
  {1992})\BibitemShut {NoStop}%
\bibitem [{\citenamefont {Aligia}\ \emph {et~al.}(2005)\citenamefont {Aligia},
  \citenamefont {Kampf},\ and\ \citenamefont {Mannhart}}]{Aligia2005}%
  \BibitemOpen
  \bibfield  {author} {\bibinfo {author} {\bibfnamefont {A.~A.}\ \bibnamefont
  {Aligia}}, \bibinfo {author} {\bibfnamefont {A.~P.}\ \bibnamefont {Kampf}},\
  and\ \bibinfo {author} {\bibfnamefont {J.}~\bibnamefont {Mannhart}},\
  }\bibfield  {title} {\bibinfo {title} {Quartet formation at $(100)/(110)$
  interfaces of $d$-wave superconductors},\ }\href
  {https://doi.org/10.1103/PhysRevLett.94.247004} {\bibfield  {journal}
  {\bibinfo  {journal} {Phys. Rev. Lett.}\ }\textbf {\bibinfo {volume} {94}},\
  \bibinfo {pages} {247004} (\bibinfo {year} {2005})}\BibitemShut {NoStop}%
\bibitem [{\citenamefont {Berg}\ \emph {et~al.}(2009)\citenamefont {Berg},
  \citenamefont {Fradkin},\ and\ \citenamefont {Kivelson}}]{Berg2009}%
  \BibitemOpen
  \bibfield  {author} {\bibinfo {author} {\bibfnamefont {E.}~\bibnamefont
  {Berg}}, \bibinfo {author} {\bibfnamefont {E.}~\bibnamefont {Fradkin}},\ and\
  \bibinfo {author} {\bibfnamefont {S.~A.}\ \bibnamefont {Kivelson}},\
  }\bibfield  {title} {\bibinfo {title} {Charge-$4e$ superconductivity from
  pair-density-wave order in certain high-temperature superconductors},\ }\href
  {https://doi.org/10.1038/nphys1389} {\bibfield  {journal} {\bibinfo
  {journal} {Nature Physics}\ }\textbf {\bibinfo {volume} {5}},\ \bibinfo
  {pages} {830} (\bibinfo {year} {2009})}\BibitemShut {NoStop}%
\bibitem [{\citenamefont {Moon}(2012)}]{Moon2012}%
  \BibitemOpen
  \bibfield  {author} {\bibinfo {author} {\bibfnamefont {E.-G.}\ \bibnamefont
  {Moon}},\ }\bibfield  {title} {\bibinfo {title} {Skyrmions with quadratic
  band touching fermions: {A} way to achieve charge $4e$ superconductivity},\
  }\href {https://doi.org/10.1103/PhysRevB.85.245123} {\bibfield  {journal}
  {\bibinfo  {journal} {Phys. Rev. B}\ }\textbf {\bibinfo {volume} {85}},\
  \bibinfo {pages} {245123} (\bibinfo {year} {2012})}\BibitemShut {NoStop}%
\bibitem [{\citenamefont {Khalaf}\ \emph {et~al.}(2022)\citenamefont {Khalaf},
  \citenamefont {Ledwith},\ and\ \citenamefont {Vishwanath}}]{Khalaf2022}%
  \BibitemOpen
  \bibfield  {author} {\bibinfo {author} {\bibfnamefont {E.}~\bibnamefont
  {Khalaf}}, \bibinfo {author} {\bibfnamefont {P.}~\bibnamefont {Ledwith}},\
  and\ \bibinfo {author} {\bibfnamefont {A.}~\bibnamefont {Vishwanath}},\
  }\bibfield  {title} {\bibinfo {title} {Symmetry constraints on
  superconductivity in twisted bilayer graphene: Fractional vortices, $4e$
  condensates, or nonunitary pairing},\ }\href
  {https://doi.org/10.1103/PhysRevB.105.224508} {\bibfield  {journal} {\bibinfo
   {journal} {Phys. Rev. B}\ }\textbf {\bibinfo {volume} {105}},\ \bibinfo
  {pages} {224508} (\bibinfo {year} {2022})}\BibitemShut {NoStop}%
\bibitem [{\citenamefont {Jiang}\ \emph {et~al.}(2017)\citenamefont {Jiang},
  \citenamefont {Li}, \citenamefont {Kivelson},\ and\ \citenamefont
  {Yao}}]{HongYao2017}%
  \BibitemOpen
  \bibfield  {author} {\bibinfo {author} {\bibfnamefont {Y.-F.}\ \bibnamefont
  {Jiang}}, \bibinfo {author} {\bibfnamefont {Z.-X.}\ \bibnamefont {Li}},
  \bibinfo {author} {\bibfnamefont {S.~A.}\ \bibnamefont {Kivelson}},\ and\
  \bibinfo {author} {\bibfnamefont {H.}~\bibnamefont {Yao}},\ }\bibfield
  {title} {\bibinfo {title} {Charge-$4e$ superconductors: A {Majorana} quantum
  {Monte Carlo} study},\ }\href {https://doi.org/10.1103/PhysRevB.95.241103}
  {\bibfield  {journal} {\bibinfo  {journal} {Phys. Rev. B}\ }\textbf {\bibinfo
  {volume} {95}},\ \bibinfo {pages} {241103} (\bibinfo {year}
  {2017})}\BibitemShut {NoStop}%
\bibitem [{\citenamefont {Gnezdilov}\ and\ \citenamefont
  {Wang}(2022)}]{Gnezdilov2021}%
  \BibitemOpen
  \bibfield  {author} {\bibinfo {author} {\bibfnamefont {N.~V.}\ \bibnamefont
  {Gnezdilov}}\ and\ \bibinfo {author} {\bibfnamefont {Y.}~\bibnamefont
  {Wang}},\ }\bibfield  {title} {\bibinfo {title} {Solvable model for a
  charge-$4e$ superconductor},\ }\href
  {https://doi.org/10.1103/PhysRevB.106.094508} {\bibfield  {journal} {\bibinfo
   {journal} {Phys. Rev. B}\ }\textbf {\bibinfo {volume} {106}},\ \bibinfo
  {pages} {094508} (\bibinfo {year} {2022})}\BibitemShut {NoStop}%
\bibitem [{\citenamefont {Ge}\ \emph {et~al.}(2022)\citenamefont {Ge},
  \citenamefont {Wang}, \citenamefont {Xing}, \citenamefont {Yin},
  \citenamefont {Lei}, \citenamefont {Wang},\ and\ \citenamefont
  {Wang}}]{Ge2022}%
  \BibitemOpen
  \bibfield  {author} {\bibinfo {author} {\bibfnamefont {J.}~\bibnamefont
  {Ge}}, \bibinfo {author} {\bibfnamefont {P.}~\bibnamefont {Wang}}, \bibinfo
  {author} {\bibfnamefont {Y.}~\bibnamefont {Xing}}, \bibinfo {author}
  {\bibfnamefont {Q.}~\bibnamefont {Yin}}, \bibinfo {author} {\bibfnamefont
  {H.}~\bibnamefont {Lei}}, \bibinfo {author} {\bibfnamefont {Z.}~\bibnamefont
  {Wang}},\ and\ \bibinfo {author} {\bibfnamefont {J.}~\bibnamefont {Wang}},\
  }\bibfield  {title} {\bibinfo {title} {Discovery of charge-$4e$ and
  charge-$6e$ superconductivity in kagome superconductor {$CsV_3Sb_5$}},\
  }\href@noop {} {\bibfield  {journal} {\bibinfo  {journal} {arXiv:2201.10352}\
  } (\bibinfo {year} {2022})}\BibitemShut {NoStop}%
\bibitem [{\citenamefont {Varma}\ and\ \citenamefont {Wang}(2023)}]{Varma2023}%
  \BibitemOpen
  \bibfield  {author} {\bibinfo {author} {\bibfnamefont {C.~M.}\ \bibnamefont
  {Varma}}\ and\ \bibinfo {author} {\bibfnamefont {Z.}~\bibnamefont {Wang}},\
  }\bibfield  {title} {\bibinfo {title} {Extended superconducting fluctuation
  region and 6e and 4e flux-quantization in a kagome compound with a normal
  state of 3q-order},\ }\href@noop {} {\bibfield  {journal} {\bibinfo
  {journal} {arXiv:2307.00448}\ } (\bibinfo {year} {2023})}\BibitemShut
  {NoStop}%
\bibitem [{\citenamefont {Nozi{\'e}res}\ and\ \citenamefont
  {Saint~James}(1982)}]{Nozieres1982}%
  \BibitemOpen
  \bibfield  {author} {\bibinfo {author} {\bibfnamefont {P.}~\bibnamefont
  {Nozi{\'e}res}}\ and\ \bibinfo {author} {\bibfnamefont {D.}~\bibnamefont
  {Saint~James}},\ }\bibfield  {title} {\bibinfo {title} {Particle vs. pair
  condensation in attractive {Bose} liquids},\ }\href
  {https://hal.science/jpa-00209488} {\bibfield  {journal} {\bibinfo  {journal}
  {Journal de Physique}\ }\textbf {\bibinfo {volume} {43}},\ \bibinfo {pages}
  {1133} (\bibinfo {year} {1982})}\BibitemShut {NoStop}%
\bibitem [{\citenamefont {Wu}(2005)}]{Wu2005}%
  \BibitemOpen
  \bibfield  {author} {\bibinfo {author} {\bibfnamefont {C.}~\bibnamefont
  {Wu}},\ }\bibfield  {title} {\bibinfo {title} {Competing orders in
  one-dimensional spin-$3/2$ fermionic systems},\ }\href
  {https://doi.org/10.1103/PhysRevLett.95.266404} {\bibfield  {journal}
  {\bibinfo  {journal} {Phys. Rev. Lett.}\ }\textbf {\bibinfo {volume} {95}},\
  \bibinfo {pages} {266404} (\bibinfo {year} {2005})}\BibitemShut {NoStop}%
\bibitem [{\citenamefont {Mukerjee}\ \emph {et~al.}(2006)\citenamefont
  {Mukerjee}, \citenamefont {Xu},\ and\ \citenamefont {Moore}}]{Moore2006}%
  \BibitemOpen
  \bibfield  {author} {\bibinfo {author} {\bibfnamefont {S.}~\bibnamefont
  {Mukerjee}}, \bibinfo {author} {\bibfnamefont {C.}~\bibnamefont {Xu}},\ and\
  \bibinfo {author} {\bibfnamefont {J.~E.}\ \bibnamefont {Moore}},\ }\bibfield
  {title} {\bibinfo {title} {Topological defects and the superfluid transition
  of the $s=1$ spinor condensate in two dimensions},\ }\href
  {https://doi.org/10.1103/PhysRevLett.97.120406} {\bibfield  {journal}
  {\bibinfo  {journal} {Phys. Rev. Lett.}\ }\textbf {\bibinfo {volume} {97}},\
  \bibinfo {pages} {120406} (\bibinfo {year} {2006})}\BibitemShut {NoStop}%
\bibitem [{\citenamefont {Herland}\ \emph {et~al.}(2010)\citenamefont
  {Herland}, \citenamefont {Babaev},\ and\ \citenamefont
  {Sudb\o{}}}]{Babaev2010}%
  \BibitemOpen
  \bibfield  {author} {\bibinfo {author} {\bibfnamefont {E.~V.}\ \bibnamefont
  {Herland}}, \bibinfo {author} {\bibfnamefont {E.}~\bibnamefont {Babaev}},\
  and\ \bibinfo {author} {\bibfnamefont {A.}~\bibnamefont {Sudb\o{}}},\
  }\bibfield  {title} {\bibinfo {title} {Phase transitions in a three
  dimensional {$U(1)\times U(1)$} lattice {London} superconductor: Metallic
  superfluid and charge-$4e$ superconducting states},\ }\href
  {https://doi.org/10.1103/PhysRevB.82.134511} {\bibfield  {journal} {\bibinfo
  {journal} {Phys. Rev. B}\ }\textbf {\bibinfo {volume} {82}},\ \bibinfo
  {pages} {134511} (\bibinfo {year} {2010})}\BibitemShut {NoStop}%
\bibitem [{\citenamefont {Radzihovsky}\ and\ \citenamefont
  {Vishwanath}(2009)}]{Radzihovsky2009}%
  \BibitemOpen
  \bibfield  {author} {\bibinfo {author} {\bibfnamefont {L.}~\bibnamefont
  {Radzihovsky}}\ and\ \bibinfo {author} {\bibfnamefont {A.}~\bibnamefont
  {Vishwanath}},\ }\bibfield  {title} {\bibinfo {title} {Quantum liquid
  crystals in an imbalanced {Fermi} gas: {Fluctuations} and fractional vortices
  in {Larkin-Ovchinnikov} states},\ }\href
  {https://doi.org/10.1103/PhysRevLett.103.010404} {\bibfield  {journal}
  {\bibinfo  {journal} {Phys. Rev. Lett.}\ }\textbf {\bibinfo {volume} {103}},\
  \bibinfo {pages} {010404} (\bibinfo {year} {2009})}\BibitemShut {NoStop}%
\bibitem [{\citenamefont {Radzihovsky}(2011)}]{Radzihovsky2011}%
  \BibitemOpen
  \bibfield  {author} {\bibinfo {author} {\bibfnamefont {L.}~\bibnamefont
  {Radzihovsky}},\ }\bibfield  {title} {\bibinfo {title} {Fluctuations and
  phase transitions in {Larkin-Ovchinnikov} liquid-crystal states of a
  population-imbalanced resonant {Fermi} gas},\ }\href
  {https://doi.org/10.1103/PhysRevA.84.023611} {\bibfield  {journal} {\bibinfo
  {journal} {Phys. Rev. A}\ }\textbf {\bibinfo {volume} {84}},\ \bibinfo
  {pages} {023611} (\bibinfo {year} {2011})}\BibitemShut {NoStop}%
\bibitem [{\citenamefont {Agterberg}\ \emph {et~al.}(2011)\citenamefont
  {Agterberg}, \citenamefont {Geracie},\ and\ \citenamefont
  {Tsunetsugu}}]{Agterberg2011}%
  \BibitemOpen
  \bibfield  {author} {\bibinfo {author} {\bibfnamefont {D.~F.}\ \bibnamefont
  {Agterberg}}, \bibinfo {author} {\bibfnamefont {M.}~\bibnamefont {Geracie}},\
  and\ \bibinfo {author} {\bibfnamefont {H.}~\bibnamefont {Tsunetsugu}},\
  }\bibfield  {title} {\bibinfo {title} {Conventional and charge-six
  superfluids from melting hexagonal {Fulde-Ferrell-Larkin-Ovchinnikov} phases
  in two dimensions},\ }\href {https://doi.org/10.1103/PhysRevB.84.014513}
  {\bibfield  {journal} {\bibinfo  {journal} {Phys. Rev. B}\ }\textbf {\bibinfo
  {volume} {84}},\ \bibinfo {pages} {014513} (\bibinfo {year}
  {2011})}\BibitemShut {NoStop}%
\bibitem [{\citenamefont {Agterberg}\ \emph {et~al.}(2020)\citenamefont
  {Agterberg}, \citenamefont {Davis}, \citenamefont {Edkins}, \citenamefont
  {Fradkin}, \citenamefont {Harlingen}, \citenamefont {Kivelson}, \citenamefont
  {Lee}, \citenamefont {Radzihovsky}, \citenamefont {Tranquada},\ and\
  \citenamefont {Wang}}]{Agterberg2020}%
  \BibitemOpen
  \bibfield  {author} {\bibinfo {author} {\bibfnamefont {D.~F.}\ \bibnamefont
  {Agterberg}}, \bibinfo {author} {\bibfnamefont {J.~C.~S.}\ \bibnamefont
  {Davis}}, \bibinfo {author} {\bibfnamefont {S.~D.}\ \bibnamefont {Edkins}},
  \bibinfo {author} {\bibfnamefont {E.}~\bibnamefont {Fradkin}}, \bibinfo
  {author} {\bibfnamefont {D.~J.~V.}\ \bibnamefont {Harlingen}}, \bibinfo
  {author} {\bibfnamefont {S.~A.}\ \bibnamefont {Kivelson}}, \bibinfo {author}
  {\bibfnamefont {P.~A.}\ \bibnamefont {Lee}}, \bibinfo {author} {\bibfnamefont
  {L.}~\bibnamefont {Radzihovsky}}, \bibinfo {author} {\bibfnamefont {J.~M.}\
  \bibnamefont {Tranquada}},\ and\ \bibinfo {author} {\bibfnamefont
  {Y.}~\bibnamefont {Wang}},\ }\bibfield  {title} {\bibinfo {title} {The
  physics of pair-density waves: Cuprate superconductors and beyond},\ }\href
  {https://doi.org/10.1146/annurev-conmatphys-031119-050711} {\bibfield
  {journal} {\bibinfo  {journal} {Annual Review of Condensed Matter Physics}\
  }\textbf {\bibinfo {volume} {11}},\ \bibinfo {pages} {231} (\bibinfo {year}
  {2020})}\BibitemShut {NoStop}%
\bibitem [{\citenamefont {Shaffer}\ \emph {et~al.}(2023)\citenamefont
  {Shaffer}, \citenamefont {Burnell},\ and\ \citenamefont
  {Fernandes}}]{Shaffer2023}%
  \BibitemOpen
  \bibfield  {author} {\bibinfo {author} {\bibfnamefont {D.}~\bibnamefont
  {Shaffer}}, \bibinfo {author} {\bibfnamefont {F.~J.}\ \bibnamefont
  {Burnell}},\ and\ \bibinfo {author} {\bibfnamefont {R.~M.}\ \bibnamefont
  {Fernandes}},\ }\bibfield  {title} {\bibinfo {title} {Weak-coupling theory of
  pair density wave instabilities in transition metal dichalcogenides},\ }\href
  {https://doi.org/10.1103/PhysRevB.107.224516} {\bibfield  {journal} {\bibinfo
   {journal} {Phys. Rev. B}\ }\textbf {\bibinfo {volume} {107}},\ \bibinfo
  {pages} {224516} (\bibinfo {year} {2023})}\BibitemShut {NoStop}%
\bibitem [{\citenamefont {Yu}(2023)}]{Yue2023}%
  \BibitemOpen
  \bibfield  {author} {\bibinfo {author} {\bibfnamefont {Y.}~\bibnamefont
  {Yu}},\ }\bibfield  {title} {\bibinfo {title} {Nondegenerate surface pair
  density wave in the kagome superconductor
  ${\mathrm{csv}}_{3}{\mathrm{sb}}_{5}$: Application to vestigial orders},\
  }\href {https://doi.org/10.1103/PhysRevB.108.054517} {\bibfield  {journal}
  {\bibinfo  {journal} {Phys. Rev. B}\ }\textbf {\bibinfo {volume} {108}},\
  \bibinfo {pages} {054517} (\bibinfo {year} {2023})}\BibitemShut {NoStop}%
\bibitem [{\citenamefont {Wu}\ and\ \citenamefont {Wang}(2023)}]{Wu_Wang2023}%
  \BibitemOpen
  \bibfield  {author} {\bibinfo {author} {\bibfnamefont {Y.-M.}\ \bibnamefont
  {Wu}}\ and\ \bibinfo {author} {\bibfnamefont {Y.}~\bibnamefont {Wang}},\
  }\bibfield  {title} {\bibinfo {title} {$ d $-wave charge-$4 e $
  superconductivity from fluctuating pair density waves},\ }\href@noop {}
  {\bibfield  {journal} {\bibinfo  {journal} {arXiv:2303.17631}\ } (\bibinfo
  {year} {2023})}\BibitemShut {NoStop}%
\bibitem [{\citenamefont {Fernandes}\ and\ \citenamefont
  {Fu}(2021)}]{Fernandes2021}%
  \BibitemOpen
  \bibfield  {author} {\bibinfo {author} {\bibfnamefont {R.~M.}\ \bibnamefont
  {Fernandes}}\ and\ \bibinfo {author} {\bibfnamefont {L.}~\bibnamefont {Fu}},\
  }\bibfield  {title} {\bibinfo {title} {Charge-$4e$ superconductivity from
  multicomponent nematic pairing: Application to twisted bilayer graphene},\
  }\href {https://doi.org/10.1103/PhysRevLett.127.047001} {\bibfield  {journal}
  {\bibinfo  {journal} {Phys. Rev. Lett.}\ }\textbf {\bibinfo {volume} {127}},\
  \bibinfo {pages} {047001} (\bibinfo {year} {2021})}\BibitemShut {NoStop}%
\bibitem [{\citenamefont {Jian}\ \emph {et~al.}(2021)\citenamefont {Jian},
  \citenamefont {Huang},\ and\ \citenamefont {Yao}}]{Jian2021}%
  \BibitemOpen
  \bibfield  {author} {\bibinfo {author} {\bibfnamefont {S.-K.}\ \bibnamefont
  {Jian}}, \bibinfo {author} {\bibfnamefont {Y.}~\bibnamefont {Huang}},\ and\
  \bibinfo {author} {\bibfnamefont {H.}~\bibnamefont {Yao}},\ }\bibfield
  {title} {\bibinfo {title} {Charge-$4e$ superconductivity from nematic
  superconductors in two and three dimensions.},\ }\href
  {https://doi.org/10.1103/PhysRevLett.127.227001} {\bibfield  {journal}
  {\bibinfo  {journal} {Phys. Rev. Lett.}\ }\textbf {\bibinfo {volume} {127}},\
  \bibinfo {pages} {227001} (\bibinfo {year} {2021})}\BibitemShut {NoStop}%
\bibitem [{\citenamefont {Zeng}\ \emph {et~al.}(2021)\citenamefont {Zeng},
  \citenamefont {Hu}, \citenamefont {Hu}, \citenamefont {You},\ and\
  \citenamefont {Wu}}]{Zeng2021}%
  \BibitemOpen
  \bibfield  {author} {\bibinfo {author} {\bibfnamefont {M.}~\bibnamefont
  {Zeng}}, \bibinfo {author} {\bibfnamefont {L.-H.}\ \bibnamefont {Hu}},
  \bibinfo {author} {\bibfnamefont {H.-Y.}\ \bibnamefont {Hu}}, \bibinfo
  {author} {\bibfnamefont {Y.-Z.}\ \bibnamefont {You}},\ and\ \bibinfo {author}
  {\bibfnamefont {C.}~\bibnamefont {Wu}},\ }\bibfield  {title} {\bibinfo
  {title} {Phase-fluctuation induced time-reversal symmetry breaking normal
  state},\ }\href@noop {} {\bibfield  {journal} {\bibinfo  {journal}
  {arXiv:2102.06158}\ } (\bibinfo {year} {2021})}\BibitemShut {NoStop}%
\bibitem [{\citenamefont {Rampp}\ \emph {et~al.}(2022)\citenamefont {Rampp},
  \citenamefont {K\"onig},\ and\ \citenamefont {Schmalian}}]{Rampp2022}%
  \BibitemOpen
  \bibfield  {author} {\bibinfo {author} {\bibfnamefont {M.~A.}\ \bibnamefont
  {Rampp}}, \bibinfo {author} {\bibfnamefont {E.~J.}\ \bibnamefont {K\"onig}},\
  and\ \bibinfo {author} {\bibfnamefont {J.}~\bibnamefont {Schmalian}},\
  }\bibfield  {title} {\bibinfo {title} {Topologically enabled
  superconductivity},\ }\href {https://doi.org/10.1103/PhysRevLett.129.077001}
  {\bibfield  {journal} {\bibinfo  {journal} {Phys. Rev. Lett.}\ }\textbf
  {\bibinfo {volume} {129}},\ \bibinfo {pages} {077001} (\bibinfo {year}
  {2022})}\BibitemShut {NoStop}%
\bibitem [{\citenamefont {Curtis}\ \emph {et~al.}(2023)\citenamefont {Curtis},
  \citenamefont {Poniatowski}, \citenamefont {Xie}, \citenamefont {Yacoby},
  \citenamefont {Demler},\ and\ \citenamefont {Narang}}]{Curtis2023}%
  \BibitemOpen
  \bibfield  {author} {\bibinfo {author} {\bibfnamefont {J.~B.}\ \bibnamefont
  {Curtis}}, \bibinfo {author} {\bibfnamefont {N.~R.}\ \bibnamefont
  {Poniatowski}}, \bibinfo {author} {\bibfnamefont {Y.}~\bibnamefont {Xie}},
  \bibinfo {author} {\bibfnamefont {A.}~\bibnamefont {Yacoby}}, \bibinfo
  {author} {\bibfnamefont {E.}~\bibnamefont {Demler}},\ and\ \bibinfo {author}
  {\bibfnamefont {P.}~\bibnamefont {Narang}},\ }\bibfield  {title} {\bibinfo
  {title} {Stabilizing fluctuating spin-triplet superconductivity in graphene
  via induced spin-orbit coupling},\ }\href
  {https://doi.org/10.1103/PhysRevLett.130.196001} {\bibfield  {journal}
  {\bibinfo  {journal} {Phys. Rev. Lett.}\ }\textbf {\bibinfo {volume} {130}},\
  \bibinfo {pages} {196001} (\bibinfo {year} {2023})}\BibitemShut {NoStop}%
\bibitem [{\citenamefont {Poduval}\ and\ \citenamefont
  {Scheurer}(2023)}]{Scheurer2023}%
  \BibitemOpen
  \bibfield  {author} {\bibinfo {author} {\bibfnamefont {P.~P.}\ \bibnamefont
  {Poduval}}\ and\ \bibinfo {author} {\bibfnamefont {M.~S.}\ \bibnamefont
  {Scheurer}},\ }\bibfield  {title} {\bibinfo {title} {Vestigial singlet
  pairing in a fluctuating magnetic triplet superconductor: Applications to
  graphene moir{\'e} systems},\ }\href@noop {} {\bibfield  {journal} {\bibinfo
  {journal} {arXiv:2301.01344}\ } (\bibinfo {year} {2023})}\BibitemShut
  {NoStop}%
\bibitem [{\citenamefont {Annett}(1990)}]{Annett1990}%
  \BibitemOpen
  \bibfield  {author} {\bibinfo {author} {\bibfnamefont {J.~F.}\ \bibnamefont
  {Annett}},\ }\bibfield  {title} {\bibinfo {title} {Symmetry of the order
  parameter for high-temperature superconductivity},\ }\href
  {https://doi.org/10.1080/00018739000101481} {\bibfield  {journal} {\bibinfo
  {journal} {Advances in Physics}\ }\textbf {\bibinfo {volume} {39}},\ \bibinfo
  {pages} {83} (\bibinfo {year} {1990})}\BibitemShut {NoStop}%
\bibitem [{\citenamefont {Sigrist}\ and\ \citenamefont
  {Ueda}(1991)}]{Sigrist1991}%
  \BibitemOpen
  \bibfield  {author} {\bibinfo {author} {\bibfnamefont {M.}~\bibnamefont
  {Sigrist}}\ and\ \bibinfo {author} {\bibfnamefont {K.}~\bibnamefont {Ueda}},\
  }\bibfield  {title} {\bibinfo {title} {Phenomenological theory of
  unconventional superconductivity},\ }\href
  {https://doi.org/10.1103/RevModPhys.63.239} {\bibfield  {journal} {\bibinfo
  {journal} {Rev. Mod. Phys.}\ }\textbf {\bibinfo {volume} {63}},\ \bibinfo
  {pages} {239} (\bibinfo {year} {1991})}\BibitemShut {NoStop}%
\bibitem [{\citenamefont {Sauls}(1994)}]{Sauls1994}%
  \BibitemOpen
  \bibfield  {author} {\bibinfo {author} {\bibfnamefont {J.}~\bibnamefont
  {Sauls}},\ }\bibfield  {title} {\bibinfo {title} {The order parameter for the
  superconducting phases of {$UPt_3$}},\ }\href
  {https://doi.org/10.1080/00018739400101475} {\bibfield  {journal} {\bibinfo
  {journal} {Advances in Physics}\ }\textbf {\bibinfo {volume} {43}},\ \bibinfo
  {pages} {113} (\bibinfo {year} {1994})}\BibitemShut {NoStop}%
\bibitem [{\citenamefont {Schemm}\ \emph {et~al.}(2014)\citenamefont {Schemm},
  \citenamefont {Gannon}, \citenamefont {Wishne}, \citenamefont {Halperin},\
  and\ \citenamefont {Kapitulnik}}]{Schemm2014}%
  \BibitemOpen
  \bibfield  {author} {\bibinfo {author} {\bibfnamefont {E.}~\bibnamefont
  {Schemm}}, \bibinfo {author} {\bibfnamefont {W.}~\bibnamefont {Gannon}},
  \bibinfo {author} {\bibfnamefont {C.}~\bibnamefont {Wishne}}, \bibinfo
  {author} {\bibfnamefont {W.}~\bibnamefont {Halperin}},\ and\ \bibinfo
  {author} {\bibfnamefont {A.}~\bibnamefont {Kapitulnik}},\ }\bibfield  {title}
  {\bibinfo {title} {Observation of broken time-reversal symmetry in the
  heavy-fermion superconductor {$UPt_3$}},\ }\href
  {https://doi.org/10.1126/science.1248552} {\bibfield  {journal} {\bibinfo
  {journal} {Science}\ }\textbf {\bibinfo {volume} {345}},\ \bibinfo {pages}
  {190} (\bibinfo {year} {2014})}\BibitemShut {NoStop}%
\bibitem [{\citenamefont {Avers}\ \emph {et~al.}(2020)\citenamefont {Avers},
  \citenamefont {Gannon}, \citenamefont {Kuhn}, \citenamefont {Halperin},
  \citenamefont {Sauls}, \citenamefont {DeBeer-Schmitt}, \citenamefont
  {Dewhurst}, \citenamefont {Gavilano}, \citenamefont {Nagy}, \citenamefont
  {Gasser} \emph {et~al.}}]{Avers2020}%
  \BibitemOpen
  \bibfield  {author} {\bibinfo {author} {\bibfnamefont {K.~E.}\ \bibnamefont
  {Avers}}, \bibinfo {author} {\bibfnamefont {W.~J.}\ \bibnamefont {Gannon}},
  \bibinfo {author} {\bibfnamefont {S.~J.}\ \bibnamefont {Kuhn}}, \bibinfo
  {author} {\bibfnamefont {W.~P.}\ \bibnamefont {Halperin}}, \bibinfo {author}
  {\bibfnamefont {J.}~\bibnamefont {Sauls}}, \bibinfo {author} {\bibfnamefont
  {L.}~\bibnamefont {DeBeer-Schmitt}}, \bibinfo {author} {\bibfnamefont
  {C.}~\bibnamefont {Dewhurst}}, \bibinfo {author} {\bibfnamefont
  {J.}~\bibnamefont {Gavilano}}, \bibinfo {author} {\bibfnamefont
  {G.}~\bibnamefont {Nagy}}, \bibinfo {author} {\bibfnamefont {U.}~\bibnamefont
  {Gasser}}, \emph {et~al.},\ }\bibfield  {title} {\bibinfo {title} {Broken
  time-reversal symmetry in the topological superconductor {$UPt_3$}},\ }\href
  {https://doi.org/10.1038/s41567-020-0822-z} {\bibfield  {journal} {\bibinfo
  {journal} {Nature Physics}\ }\textbf {\bibinfo {volume} {16}},\ \bibinfo
  {pages} {531} (\bibinfo {year} {2020})}\BibitemShut {NoStop}%
\bibitem [{\citenamefont {Cao}\ \emph {et~al.}(2021)\citenamefont {Cao},
  \citenamefont {Rodan-Legrain}, \citenamefont {Park}, \citenamefont {Yuan},
  \citenamefont {Watanabe}, \citenamefont {Taniguchi}, \citenamefont
  {Fernandes}, \citenamefont {Fu},\ and\ \citenamefont
  {Jarillo-Herrero}}]{Cao2021}%
  \BibitemOpen
  \bibfield  {author} {\bibinfo {author} {\bibfnamefont {Y.}~\bibnamefont
  {Cao}}, \bibinfo {author} {\bibfnamefont {D.}~\bibnamefont {Rodan-Legrain}},
  \bibinfo {author} {\bibfnamefont {J.~M.}\ \bibnamefont {Park}}, \bibinfo
  {author} {\bibfnamefont {N.~F.}\ \bibnamefont {Yuan}}, \bibinfo {author}
  {\bibfnamefont {K.}~\bibnamefont {Watanabe}}, \bibinfo {author}
  {\bibfnamefont {T.}~\bibnamefont {Taniguchi}}, \bibinfo {author}
  {\bibfnamefont {R.~M.}\ \bibnamefont {Fernandes}}, \bibinfo {author}
  {\bibfnamefont {L.}~\bibnamefont {Fu}},\ and\ \bibinfo {author}
  {\bibfnamefont {P.}~\bibnamefont {Jarillo-Herrero}},\ }\bibfield  {title}
  {\bibinfo {title} {Nematicity and competing orders in superconducting
  magic-angle graphene},\ }\href {https://doi.org/10.1126/science.abc2836}
  {\bibfield  {journal} {\bibinfo  {journal} {Science}\ }\textbf {\bibinfo
  {volume} {372}},\ \bibinfo {pages} {264} (\bibinfo {year}
  {2021})}\BibitemShut {NoStop}%
\bibitem [{\citenamefont {Matano}\ \emph {et~al.}(2016)\citenamefont {Matano},
  \citenamefont {Kriener}, \citenamefont {Segawa}, \citenamefont {Ando},\ and\
  \citenamefont {Zheng}}]{Matano2015}%
  \BibitemOpen
  \bibfield  {author} {\bibinfo {author} {\bibfnamefont {K.}~\bibnamefont
  {Matano}}, \bibinfo {author} {\bibfnamefont {M.}~\bibnamefont {Kriener}},
  \bibinfo {author} {\bibfnamefont {K.}~\bibnamefont {Segawa}}, \bibinfo
  {author} {\bibfnamefont {Y.}~\bibnamefont {Ando}},\ and\ \bibinfo {author}
  {\bibfnamefont {G.-q.}\ \bibnamefont {Zheng}},\ }\bibfield  {title} {\bibinfo
  {title} {Spin-rotation symmetry breaking in the superconducting state of
  {$Cu_xBi_2Se_3$}},\ }\href {https://doi.org/10.1038/nphys3781} {\bibfield
  {journal} {\bibinfo  {journal} {Nature Physics}\ }\textbf {\bibinfo {volume}
  {12}},\ \bibinfo {pages} {852} (\bibinfo {year} {2016})}\BibitemShut
  {NoStop}%
\bibitem [{\citenamefont {Pan}\ \emph {et~al.}(2016)\citenamefont {Pan},
  \citenamefont {Nikitin}, \citenamefont {Araizi}, \citenamefont {Huang},
  \citenamefont {Matsushita}, \citenamefont {Naka},\ and\ \citenamefont
  {De~Visser}}]{Pan2016}%
  \BibitemOpen
  \bibfield  {author} {\bibinfo {author} {\bibfnamefont {Y.}~\bibnamefont
  {Pan}}, \bibinfo {author} {\bibfnamefont {A.}~\bibnamefont {Nikitin}},
  \bibinfo {author} {\bibfnamefont {G.}~\bibnamefont {Araizi}}, \bibinfo
  {author} {\bibfnamefont {Y.}~\bibnamefont {Huang}}, \bibinfo {author}
  {\bibfnamefont {Y.}~\bibnamefont {Matsushita}}, \bibinfo {author}
  {\bibfnamefont {T.}~\bibnamefont {Naka}},\ and\ \bibinfo {author}
  {\bibfnamefont {A.}~\bibnamefont {De~Visser}},\ }\bibfield  {title} {\bibinfo
  {title} {Rotational symmetry breaking in the topological superconductor
  {$Sr_x Bi_2 Se_3$} probed by upper-critical field experiments},\ }\href
  {https://doi.org/10.1038/srep28632 10.1038/srep28632} {\bibfield  {journal}
  {\bibinfo  {journal} {Scientific reports}\ }\textbf {\bibinfo {volume} {6}},\
  \bibinfo {pages} {28632} (\bibinfo {year} {2016})}\BibitemShut {NoStop}%
\bibitem [{\citenamefont {Asaba}\ \emph {et~al.}(2017)\citenamefont {Asaba},
  \citenamefont {Lawson}, \citenamefont {Tinsman}, \citenamefont {Chen},
  \citenamefont {Corbae}, \citenamefont {Li}, \citenamefont {Qiu},
  \citenamefont {Hor}, \citenamefont {Fu},\ and\ \citenamefont
  {Li}}]{Asaba2016}%
  \BibitemOpen
  \bibfield  {author} {\bibinfo {author} {\bibfnamefont {T.}~\bibnamefont
  {Asaba}}, \bibinfo {author} {\bibfnamefont {B.~J.}\ \bibnamefont {Lawson}},
  \bibinfo {author} {\bibfnamefont {C.}~\bibnamefont {Tinsman}}, \bibinfo
  {author} {\bibfnamefont {L.}~\bibnamefont {Chen}}, \bibinfo {author}
  {\bibfnamefont {P.}~\bibnamefont {Corbae}}, \bibinfo {author} {\bibfnamefont
  {G.}~\bibnamefont {Li}}, \bibinfo {author} {\bibfnamefont {Y.}~\bibnamefont
  {Qiu}}, \bibinfo {author} {\bibfnamefont {Y.~S.}\ \bibnamefont {Hor}},
  \bibinfo {author} {\bibfnamefont {L.}~\bibnamefont {Fu}},\ and\ \bibinfo
  {author} {\bibfnamefont {L.}~\bibnamefont {Li}},\ }\bibfield  {title}
  {\bibinfo {title} {Rotational symmetry breaking in a trigonal superconductor
  {$Nb$}-doped {$Bi_2Se_3$}},\ }\href
  {https://doi.org/10.1103/PhysRevX.7.011009} {\bibfield  {journal} {\bibinfo
  {journal} {Phys. Rev. X}\ }\textbf {\bibinfo {volume} {7}},\ \bibinfo {pages}
  {011009} (\bibinfo {year} {2017})}\BibitemShut {NoStop}%
\bibitem [{\citenamefont {Hecker}\ \emph {et~al.}(2023)\citenamefont {Hecker},
  \citenamefont {Willa}, \citenamefont {Schmalian},\ and\ \citenamefont
  {Fernandes}}]{Hecker2023}%
  \BibitemOpen
  \bibfield  {author} {\bibinfo {author} {\bibfnamefont {M.}~\bibnamefont
  {Hecker}}, \bibinfo {author} {\bibfnamefont {R.}~\bibnamefont {Willa}},
  \bibinfo {author} {\bibfnamefont {J.}~\bibnamefont {Schmalian}},\ and\
  \bibinfo {author} {\bibfnamefont {R.~M.}\ \bibnamefont {Fernandes}},\
  }\bibfield  {title} {\bibinfo {title} {Cascade of vestigial orders in
  two-component superconductors: Nematic, ferromagnetic, $s$-wave charge-$4e$,
  and $d$-wave charge-$4e$ states},\ }\href
  {https://doi.org/10.1103/PhysRevB.107.224503} {\bibfield  {journal} {\bibinfo
   {journal} {Phys. Rev. B}\ }\textbf {\bibinfo {volume} {107}},\ \bibinfo
  {pages} {224503} (\bibinfo {year} {2023})}\BibitemShut {NoStop}%
\bibitem [{\citenamefont {Fradkin}\ \emph {et~al.}(2015)\citenamefont
  {Fradkin}, \citenamefont {Kivelson},\ and\ \citenamefont
  {Tranquada}}]{Fradkin2015}%
  \BibitemOpen
  \bibfield  {author} {\bibinfo {author} {\bibfnamefont {E.}~\bibnamefont
  {Fradkin}}, \bibinfo {author} {\bibfnamefont {S.~A.}\ \bibnamefont
  {Kivelson}},\ and\ \bibinfo {author} {\bibfnamefont {J.~M.}\ \bibnamefont
  {Tranquada}},\ }\bibfield  {title} {\bibinfo {title} {Colloquium: Theory of
  intertwined orders in high temperature superconductors},\ }\href
  {https://doi.org/10.1103/RevModPhys.87.457} {\bibfield  {journal} {\bibinfo
  {journal} {Rev. Mod. Phys.}\ }\textbf {\bibinfo {volume} {87}},\ \bibinfo
  {pages} {457} (\bibinfo {year} {2015})}\BibitemShut {NoStop}%
\bibitem [{\citenamefont {Fernandes}\ \emph {et~al.}(2019)\citenamefont
  {Fernandes}, \citenamefont {Orth},\ and\ \citenamefont
  {Schmalian}}]{Fernandes2019}%
  \BibitemOpen
  \bibfield  {author} {\bibinfo {author} {\bibfnamefont {R.~M.}\ \bibnamefont
  {Fernandes}}, \bibinfo {author} {\bibfnamefont {P.~P.}\ \bibnamefont
  {Orth}},\ and\ \bibinfo {author} {\bibfnamefont {J.}~\bibnamefont
  {Schmalian}},\ }\bibfield  {title} {\bibinfo {title} {Intertwined vestigial
  order in quantum materials: Nematicity and beyond},\ }\href
  {https://doi.org/10.1146/annurev-conmatphys-031218-013200} {\bibfield
  {journal} {\bibinfo  {journal} {Annual Review of Condensed Matter Physics}\
  }\textbf {\bibinfo {volume} {10}},\ \bibinfo {pages} {133} (\bibinfo {year}
  {2019})}\BibitemShut {NoStop}%
\bibitem [{\citenamefont {Gali}\ and\ \citenamefont
  {Fernandes}(2022)}]{Gali2022}%
  \BibitemOpen
  \bibfield  {author} {\bibinfo {author} {\bibfnamefont {V.}~\bibnamefont
  {Gali}}\ and\ \bibinfo {author} {\bibfnamefont {R.~M.}\ \bibnamefont
  {Fernandes}},\ }\bibfield  {title} {\bibinfo {title} {Role of electromagnetic
  gauge-field fluctuations in the selection between chiral and nematic
  superconductivity},\ }\href {https://doi.org/10.1103/PhysRevB.106.094509}
  {\bibfield  {journal} {\bibinfo  {journal} {Phys. Rev. B}\ }\textbf {\bibinfo
  {volume} {106}},\ \bibinfo {pages} {094509} (\bibinfo {year}
  {2022})}\BibitemShut {NoStop}%
\bibitem [{\citenamefont {Hecker}\ and\ \citenamefont
  {Schmalian}(2018)}]{Hecker2018}%
  \BibitemOpen
  \bibfield  {author} {\bibinfo {author} {\bibfnamefont {M.}~\bibnamefont
  {Hecker}}\ and\ \bibinfo {author} {\bibfnamefont {J.}~\bibnamefont
  {Schmalian}},\ }\bibfield  {title} {\bibinfo {title} {Vestigial nematic order
  and superconductivity in the doped topological insulator {$Cu_x Bi_2
  Se_3$}},\ }\href {https://doi.org/10.1038/s41535-018-0098-z} {\bibfield
  {journal} {\bibinfo  {journal} {npj Quantum Materials}\ }\textbf {\bibinfo
  {volume} {3}},\ \bibinfo {pages} {1} (\bibinfo {year} {2018})}\BibitemShut
  {NoStop}%
\bibitem [{\citenamefont {Willa}(2020)}]{Willa2020}%
  \BibitemOpen
  \bibfield  {author} {\bibinfo {author} {\bibfnamefont {R.}~\bibnamefont
  {Willa}},\ }\bibfield  {title} {\bibinfo {title} {Symmetry-mixed bound-state
  order},\ }\href {https://doi.org/10.1103/PhysRevB.102.180503} {\bibfield
  {journal} {\bibinfo  {journal} {Phys. Rev. B}\ }\textbf {\bibinfo {volume}
  {102}},\ \bibinfo {pages} {180503} (\bibinfo {year} {2020})}\BibitemShut
  {NoStop}%
\bibitem [{\citenamefont {Venderbos}\ and\ \citenamefont
  {Fernandes}(2018)}]{Venderbos_Fernandes}%
  \BibitemOpen
  \bibfield  {author} {\bibinfo {author} {\bibfnamefont {J.~W.~F.}\
  \bibnamefont {Venderbos}}\ and\ \bibinfo {author} {\bibfnamefont {R.~M.}\
  \bibnamefont {Fernandes}},\ }\bibfield  {title} {\bibinfo {title}
  {Correlations and electronic order in a two-orbital honeycomb lattice model
  for twisted bilayer graphene},\ }\href
  {https://doi.org/10.1103/PhysRevB.98.245103} {\bibfield  {journal} {\bibinfo
  {journal} {Phys. Rev. B}\ }\textbf {\bibinfo {volume} {98}},\ \bibinfo
  {pages} {245103} (\bibinfo {year} {2018})}\BibitemShut {NoStop}%
\bibitem [{\citenamefont {Scheurer}\ and\ \citenamefont
  {Samajdar}(2020)}]{Scheurer2020}%
  \BibitemOpen
  \bibfield  {author} {\bibinfo {author} {\bibfnamefont {M.~S.}\ \bibnamefont
  {Scheurer}}\ and\ \bibinfo {author} {\bibfnamefont {R.}~\bibnamefont
  {Samajdar}},\ }\bibfield  {title} {\bibinfo {title} {Pairing in
  graphene-based moir\'e superlattices},\ }\href
  {https://doi.org/10.1103/PhysRevResearch.2.033062} {\bibfield  {journal}
  {\bibinfo  {journal} {Phys. Rev. Research}\ }\textbf {\bibinfo {volume}
  {2}},\ \bibinfo {pages} {033062} (\bibinfo {year} {2020})}\BibitemShut
  {NoStop}%
\bibitem [{\citenamefont {Wang}\ \emph {et~al.}(2021)\citenamefont {Wang},
  \citenamefont {Kang},\ and\ \citenamefont {Fernandes}}]{Wang_Kang2021}%
  \BibitemOpen
  \bibfield  {author} {\bibinfo {author} {\bibfnamefont {Y.}~\bibnamefont
  {Wang}}, \bibinfo {author} {\bibfnamefont {J.}~\bibnamefont {Kang}},\ and\
  \bibinfo {author} {\bibfnamefont {R.~M.}\ \bibnamefont {Fernandes}},\
  }\bibfield  {title} {\bibinfo {title} {Topological and nematic
  superconductivity mediated by ferro-su(4) fluctuations in twisted bilayer
  graphene},\ }\href {https://doi.org/10.1103/PhysRevB.103.024506} {\bibfield
  {journal} {\bibinfo  {journal} {Phys. Rev. B}\ }\textbf {\bibinfo {volume}
  {103}},\ \bibinfo {pages} {024506} (\bibinfo {year} {2021})}\BibitemShut
  {NoStop}%
\bibitem [{\citenamefont {Lake}\ \emph {et~al.}(2022)\citenamefont {Lake},
  \citenamefont {Patri},\ and\ \citenamefont {Senthil}}]{Lake2022}%
  \BibitemOpen
  \bibfield  {author} {\bibinfo {author} {\bibfnamefont {E.}~\bibnamefont
  {Lake}}, \bibinfo {author} {\bibfnamefont {A.~S.}\ \bibnamefont {Patri}},\
  and\ \bibinfo {author} {\bibfnamefont {T.}~\bibnamefont {Senthil}},\
  }\bibfield  {title} {\bibinfo {title} {Pairing symmetry of twisted bilayer
  graphene: A phenomenological synthesis},\ }\href
  {https://doi.org/10.1103/PhysRevB.106.104506} {\bibfield  {journal} {\bibinfo
   {journal} {Phys. Rev. B}\ }\textbf {\bibinfo {volume} {106}},\ \bibinfo
  {pages} {104506} (\bibinfo {year} {2022})}\BibitemShut {NoStop}%
\bibitem [{\citenamefont {Dong}\ \emph {et~al.}(2023)\citenamefont {Dong},
  \citenamefont {Chubukov},\ and\ \citenamefont {Levitov}}]{Levitov_Chubukov}%
  \BibitemOpen
  \bibfield  {author} {\bibinfo {author} {\bibfnamefont {Z.}~\bibnamefont
  {Dong}}, \bibinfo {author} {\bibfnamefont {A.~V.}\ \bibnamefont {Chubukov}},\
  and\ \bibinfo {author} {\bibfnamefont {L.}~\bibnamefont {Levitov}},\
  }\bibfield  {title} {\bibinfo {title} {Transformer spin-triplet
  superconductivity at the onset of isospin order in bilayer graphene},\ }\href
  {https://doi.org/10.1103/PhysRevB.107.174512} {\bibfield  {journal} {\bibinfo
   {journal} {Phys. Rev. B}\ }\textbf {\bibinfo {volume} {107}},\ \bibinfo
  {pages} {174512} (\bibinfo {year} {2023})}\BibitemShut {NoStop}%
\bibitem [{\citenamefont {Volovik}\ and\ \citenamefont
  {Gor'kov}(1985)}]{Volovik1985}%
  \BibitemOpen
  \bibfield  {author} {\bibinfo {author} {\bibfnamefont {G.}~\bibnamefont
  {Volovik}}\ and\ \bibinfo {author} {\bibfnamefont {L.}~\bibnamefont
  {Gor'kov}},\ }\bibfield  {title} {\bibinfo {title} {Superconducting classes
  in heavy-fermion systems},\ }\href@noop {} {\bibfield  {journal} {\bibinfo
  {journal} {Zh. Eksp. Teor. Fiz}\ }\textbf {\bibinfo {volume} {88}},\ \bibinfo
  {pages} {1412} (\bibinfo {year} {1985})}\BibitemShut {NoStop}%
\bibitem [{\citenamefont {Blagoeva}\ \emph {et~al.}(1990)\citenamefont
  {Blagoeva}, \citenamefont {Busiello}, \citenamefont {De~Cesare},
  \citenamefont {Millev}, \citenamefont {Rabuffo},\ and\ \citenamefont
  {Uzunov}}]{Uzunov1990}%
  \BibitemOpen
  \bibfield  {author} {\bibinfo {author} {\bibfnamefont {E.~J.}\ \bibnamefont
  {Blagoeva}}, \bibinfo {author} {\bibfnamefont {G.}~\bibnamefont {Busiello}},
  \bibinfo {author} {\bibfnamefont {L.}~\bibnamefont {De~Cesare}}, \bibinfo
  {author} {\bibfnamefont {Y.~T.}\ \bibnamefont {Millev}}, \bibinfo {author}
  {\bibfnamefont {I.}~\bibnamefont {Rabuffo}},\ and\ \bibinfo {author}
  {\bibfnamefont {D.~I.}\ \bibnamefont {Uzunov}},\ }\bibfield  {title}
  {\bibinfo {title} {Fluctuation-induced first-order transitions in
  unconventional superconductors},\ }\href
  {https://doi.org/10.1103/PhysRevB.42.6124} {\bibfield  {journal} {\bibinfo
  {journal} {Phys. Rev. B}\ }\textbf {\bibinfo {volume} {42}},\ \bibinfo
  {pages} {6124} (\bibinfo {year} {1990})}\BibitemShut {NoStop}%
\bibitem [{\citenamefont {Blume}\ and\ \citenamefont
  {Hsieh}(1969)}]{Blume1969}%
  \BibitemOpen
  \bibfield  {author} {\bibinfo {author} {\bibfnamefont {M.}~\bibnamefont
  {Blume}}\ and\ \bibinfo {author} {\bibfnamefont {Y.}~\bibnamefont {Hsieh}},\
  }\bibfield  {title} {\bibinfo {title} {Biquadratic exchange and quadrupolar
  ordering},\ }\href {https://doi.org/10.1063/1.1657616} {\bibfield  {journal}
  {\bibinfo  {journal} {J. Appl. Phys.}\ }\textbf {\bibinfo {volume} {40}},\
  \bibinfo {pages} {1249} (\bibinfo {year} {1969})}\BibitemShut {NoStop}%
\bibitem [{\citenamefont {Andreev}\ and\ \citenamefont
  {Grishchuk}(1984)}]{Andreev1984}%
  \BibitemOpen
  \bibfield  {author} {\bibinfo {author} {\bibfnamefont {A.}~\bibnamefont
  {Andreev}}\ and\ \bibinfo {author} {\bibfnamefont {I.}~\bibnamefont
  {Grishchuk}},\ }\bibfield  {title} {\bibinfo {title} {Spin nematics},\
  }\href@noop {} {\bibfield  {journal} {\bibinfo  {journal} {Sov. Phys. JETP}\
  }\textbf {\bibinfo {volume} {60}},\ \bibinfo {pages} {267} (\bibinfo {year}
  {1984})}\BibitemShut {NoStop}%
\bibitem [{\citenamefont {Haenel}\ \emph {et~al.}(2022)\citenamefont {Haenel},
  \citenamefont {Tummuru},\ and\ \citenamefont {Franz}}]{Franz2022}%
  \BibitemOpen
  \bibfield  {author} {\bibinfo {author} {\bibfnamefont {R.}~\bibnamefont
  {Haenel}}, \bibinfo {author} {\bibfnamefont {T.}~\bibnamefont {Tummuru}},\
  and\ \bibinfo {author} {\bibfnamefont {M.}~\bibnamefont {Franz}},\ }\bibfield
   {title} {\bibinfo {title} {Incoherent tunneling and topological
  superconductivity in twisted cuprate bilayers},\ }\href
  {https://doi.org/10.1103/PhysRevB.106.104505} {\bibfield  {journal} {\bibinfo
   {journal} {Phys. Rev. B}\ }\textbf {\bibinfo {volume} {106}},\ \bibinfo
  {pages} {104505} (\bibinfo {year} {2022})}\BibitemShut {NoStop}%
\bibitem [{\citenamefont {Moshe}\ and\ \citenamefont
  {Zinn-Justin}(2003)}]{Moshe2003}%
  \BibitemOpen
  \bibfield  {author} {\bibinfo {author} {\bibfnamefont {M.}~\bibnamefont
  {Moshe}}\ and\ \bibinfo {author} {\bibfnamefont {J.}~\bibnamefont
  {Zinn-Justin}},\ }\bibfield  {title} {\bibinfo {title} {Quantum field theory
  in the large {N} limit: a review},\ }\href
  {https://doi.org/https://doi.org/10.1016/S0370-1573(03)00263-1} {\bibfield
  {journal} {\bibinfo  {journal} {Physics Reports}\ }\textbf {\bibinfo {volume}
  {385}},\ \bibinfo {pages} {69} (\bibinfo {year} {2003})}\BibitemShut
  {NoStop}%
\bibitem [{\citenamefont {Fischer}\ and\ \citenamefont
  {Berg}(2016)}]{Fischer2016}%
  \BibitemOpen
  \bibfield  {author} {\bibinfo {author} {\bibfnamefont {M.~H.}\ \bibnamefont
  {Fischer}}\ and\ \bibinfo {author} {\bibfnamefont {E.}~\bibnamefont {Berg}},\
  }\bibfield  {title} {\bibinfo {title} {Fluctuation and strain effects in a
  chiral $p$-wave superconductor},\ }\href
  {https://doi.org/10.1103/PhysRevB.93.054501} {\bibfield  {journal} {\bibinfo
  {journal} {Phys. Rev. B}\ }\textbf {\bibinfo {volume} {93}},\ \bibinfo
  {pages} {054501} (\bibinfo {year} {2016})}\BibitemShut {NoStop}%
\bibitem [{\citenamefont {Nie}\ \emph {et~al.}(2017)\citenamefont {Nie},
  \citenamefont {Maharaj}, \citenamefont {Fradkin},\ and\ \citenamefont
  {Kivelson}}]{Nie2017}%
  \BibitemOpen
  \bibfield  {author} {\bibinfo {author} {\bibfnamefont {L.}~\bibnamefont
  {Nie}}, \bibinfo {author} {\bibfnamefont {A.~V.}\ \bibnamefont {Maharaj}},
  \bibinfo {author} {\bibfnamefont {E.}~\bibnamefont {Fradkin}},\ and\ \bibinfo
  {author} {\bibfnamefont {S.~A.}\ \bibnamefont {Kivelson}},\ }\bibfield
  {title} {\bibinfo {title} {Vestigial nematicity from spin and/or charge order
  in the cuprates},\ }\href {https://doi.org/10.1103/PhysRevB.96.085142}
  {\bibfield  {journal} {\bibinfo  {journal} {Phys. Rev. B}\ }\textbf {\bibinfo
  {volume} {96}},\ \bibinfo {pages} {085142} (\bibinfo {year}
  {2017})}\BibitemShut {NoStop}%
\bibitem [{\citenamefont {Fu}(2014)}]{Fu2014}%
  \BibitemOpen
  \bibfield  {author} {\bibinfo {author} {\bibfnamefont {L.}~\bibnamefont
  {Fu}},\ }\bibfield  {title} {\bibinfo {title} {Odd-parity topological
  superconductor with nematic order: Application to {$Cu_x Bi_2 Se_3$}},\
  }\href {https://doi.org/10.1103/PhysRevB.90.100509} {\bibfield  {journal}
  {\bibinfo  {journal} {Phys. Rev. B}\ }\textbf {\bibinfo {volume} {90}},\
  \bibinfo {pages} {100509} (\bibinfo {year} {2014})}\BibitemShut {NoStop}%
\bibitem [{\citenamefont {Kozii}\ \emph {et~al.}(2019)\citenamefont {Kozii},
  \citenamefont {Isobe}, \citenamefont {Venderbos},\ and\ \citenamefont
  {Fu}}]{Kozii2019}%
  \BibitemOpen
  \bibfield  {author} {\bibinfo {author} {\bibfnamefont {V.}~\bibnamefont
  {Kozii}}, \bibinfo {author} {\bibfnamefont {H.}~\bibnamefont {Isobe}},
  \bibinfo {author} {\bibfnamefont {J.~W.~F.}\ \bibnamefont {Venderbos}},\ and\
  \bibinfo {author} {\bibfnamefont {L.}~\bibnamefont {Fu}},\ }\bibfield
  {title} {\bibinfo {title} {Nematic superconductivity stabilized by density
  wave fluctuations: {Possible} application to twisted bilayer graphene},\
  }\href {https://doi.org/10.1103/PhysRevB.99.144507} {\bibfield  {journal}
  {\bibinfo  {journal} {Phys. Rev. B}\ }\textbf {\bibinfo {volume} {99}},\
  \bibinfo {pages} {144507} (\bibinfo {year} {2019})}\BibitemShut {NoStop}%
\bibitem [{\citenamefont {Chichinadze}\ \emph {et~al.}(2020)\citenamefont
  {Chichinadze}, \citenamefont {Classen},\ and\ \citenamefont
  {Chubukov}}]{Chichinadze2020}%
  \BibitemOpen
  \bibfield  {author} {\bibinfo {author} {\bibfnamefont {D.~V.}\ \bibnamefont
  {Chichinadze}}, \bibinfo {author} {\bibfnamefont {L.}~\bibnamefont
  {Classen}},\ and\ \bibinfo {author} {\bibfnamefont {A.~V.}\ \bibnamefont
  {Chubukov}},\ }\bibfield  {title} {\bibinfo {title} {Nematic
  superconductivity in twisted bilayer graphene},\ }\href
  {https://doi.org/10.1103/PhysRevB.101.224513} {\bibfield  {journal} {\bibinfo
   {journal} {Phys. Rev. B}\ }\textbf {\bibinfo {volume} {101}},\ \bibinfo
  {pages} {224513} (\bibinfo {year} {2020})}\BibitemShut {NoStop}%
\bibitem [{\citenamefont {Sun}\ \emph {et~al.}(2019)\citenamefont {Sun},
  \citenamefont {Kittaka}, \citenamefont {Sakakibara}, \citenamefont {Machida},
  \citenamefont {Wang}, \citenamefont {Wen}, \citenamefont {Xing},
  \citenamefont {Shi},\ and\ \citenamefont {Tamegai}}]{Tamegai2019}%
  \BibitemOpen
  \bibfield  {author} {\bibinfo {author} {\bibfnamefont {Y.}~\bibnamefont
  {Sun}}, \bibinfo {author} {\bibfnamefont {S.}~\bibnamefont {Kittaka}},
  \bibinfo {author} {\bibfnamefont {T.}~\bibnamefont {Sakakibara}}, \bibinfo
  {author} {\bibfnamefont {K.}~\bibnamefont {Machida}}, \bibinfo {author}
  {\bibfnamefont {J.}~\bibnamefont {Wang}}, \bibinfo {author} {\bibfnamefont
  {J.}~\bibnamefont {Wen}}, \bibinfo {author} {\bibfnamefont {X.}~\bibnamefont
  {Xing}}, \bibinfo {author} {\bibfnamefont {Z.}~\bibnamefont {Shi}},\ and\
  \bibinfo {author} {\bibfnamefont {T.}~\bibnamefont {Tamegai}},\ }\bibfield
  {title} {\bibinfo {title} {Quasiparticle evidence for the nematic state above
  {${T}_{\mathrm{c}}$} in {$ Sr_x Bi_2 Se_3$}},\ }\href
  {https://doi.org/10.1103/PhysRevLett.123.027002} {\bibfield  {journal}
  {\bibinfo  {journal} {Phys. Rev. Lett.}\ }\textbf {\bibinfo {volume} {123}},\
  \bibinfo {pages} {027002} (\bibinfo {year} {2019})}\BibitemShut {NoStop}%
\bibitem [{\citenamefont {Cho}\ \emph {et~al.}(2020)\citenamefont {Cho},
  \citenamefont {Shen}, \citenamefont {Lyu}, \citenamefont {Atanov},
  \citenamefont {Chen}, \citenamefont {Lee}, \citenamefont {San~Hor},
  \citenamefont {Gawryluk}, \citenamefont {Pomjakushina}, \citenamefont
  {Bartkowiak}, \citenamefont {Hecker}, \citenamefont {Schmalian},\ and\
  \citenamefont {Lortz}}]{Cho2019}%
  \BibitemOpen
  \bibfield  {author} {\bibinfo {author} {\bibfnamefont {C.-w.}\ \bibnamefont
  {Cho}}, \bibinfo {author} {\bibfnamefont {J.}~\bibnamefont {Shen}}, \bibinfo
  {author} {\bibfnamefont {J.}~\bibnamefont {Lyu}}, \bibinfo {author}
  {\bibfnamefont {O.}~\bibnamefont {Atanov}}, \bibinfo {author} {\bibfnamefont
  {Q.}~\bibnamefont {Chen}}, \bibinfo {author} {\bibfnamefont {S.~H.}\
  \bibnamefont {Lee}}, \bibinfo {author} {\bibfnamefont {Y.}~\bibnamefont
  {San~Hor}}, \bibinfo {author} {\bibfnamefont {D.~J.}\ \bibnamefont
  {Gawryluk}}, \bibinfo {author} {\bibfnamefont {E.}~\bibnamefont
  {Pomjakushina}}, \bibinfo {author} {\bibfnamefont {M.}~\bibnamefont
  {Bartkowiak}}, \bibinfo {author} {\bibfnamefont {M.}~\bibnamefont {Hecker}},
  \bibinfo {author} {\bibfnamefont {J.}~\bibnamefont {Schmalian}},\ and\
  \bibinfo {author} {\bibfnamefont {R.}~\bibnamefont {Lortz}},\ }\bibfield
  {title} {\bibinfo {title} {{$Z_3$}-vestigial nematic order due to
  superconducting fluctuations in the doped topological insulator
  {$Nb_xBi_2Se_3$} and {$Cu_xBi_2Se_3$}},\ }\href
  {https://doi.org/10.1038/s41467-020-16871-9} {\bibfield  {journal} {\bibinfo
  {journal} {Nature communications}\ }\textbf {\bibinfo {volume} {11}},\
  \bibinfo {pages} {1} (\bibinfo {year} {2020})}\BibitemShut {NoStop}%
\bibitem [{\citenamefont {Ghosh}\ \emph {et~al.}(2020)\citenamefont {Ghosh},
  \citenamefont {Smidman}, \citenamefont {Shang}, \citenamefont {Annett},
  \citenamefont {Hillier}, \citenamefont {Quintanilla},\ and\ \citenamefont
  {Yuan}}]{Ghosh2020}%
  \BibitemOpen
  \bibfield  {author} {\bibinfo {author} {\bibfnamefont {S.~K.}\ \bibnamefont
  {Ghosh}}, \bibinfo {author} {\bibfnamefont {M.}~\bibnamefont {Smidman}},
  \bibinfo {author} {\bibfnamefont {T.}~\bibnamefont {Shang}}, \bibinfo
  {author} {\bibfnamefont {J.~F.}\ \bibnamefont {Annett}}, \bibinfo {author}
  {\bibfnamefont {A.~D.}\ \bibnamefont {Hillier}}, \bibinfo {author}
  {\bibfnamefont {J.}~\bibnamefont {Quintanilla}},\ and\ \bibinfo {author}
  {\bibfnamefont {H.}~\bibnamefont {Yuan}},\ }\bibfield  {title} {\bibinfo
  {title} {Recent progress on superconductors with time-reversal symmetry
  breaking},\ }\href {https://doi.org/10.1088/1361-648X/abaa06} {\bibfield
  {journal} {\bibinfo  {journal} {Journal of Physics: Condensed Matter}\
  }\textbf {\bibinfo {volume} {33}},\ \bibinfo {pages} {033001} (\bibinfo
  {year} {2020})}\BibitemShut {NoStop}%
\bibitem [{\citenamefont {Naaman}\ \emph {et~al.}(2001)\citenamefont {Naaman},
  \citenamefont {Teizer},\ and\ \citenamefont {Dynes}}]{Naaman2001}%
  \BibitemOpen
  \bibfield  {author} {\bibinfo {author} {\bibfnamefont {O.}~\bibnamefont
  {Naaman}}, \bibinfo {author} {\bibfnamefont {W.}~\bibnamefont {Teizer}},\
  and\ \bibinfo {author} {\bibfnamefont {R.~C.}\ \bibnamefont {Dynes}},\
  }\bibfield  {title} {\bibinfo {title} {Fluctuation dominated josephson
  tunneling with a scanning tunneling microscope},\ }\href
  {https://doi.org/10.1103/PhysRevLett.87.097004} {\bibfield  {journal}
  {\bibinfo  {journal} {Phys. Rev. Lett.}\ }\textbf {\bibinfo {volume} {87}},\
  \bibinfo {pages} {097004} (\bibinfo {year} {2001})}\BibitemShut {NoStop}%
\bibitem [{\citenamefont {Bastiaans}\ \emph {et~al.}(2019)\citenamefont
  {Bastiaans}, \citenamefont {Cho}, \citenamefont {Chatzopoulos}, \citenamefont
  {Leeuwenhoek}, \citenamefont {Koks},\ and\ \citenamefont
  {Allan}}]{Allan2019}%
  \BibitemOpen
  \bibfield  {author} {\bibinfo {author} {\bibfnamefont {K.~M.}\ \bibnamefont
  {Bastiaans}}, \bibinfo {author} {\bibfnamefont {D.}~\bibnamefont {Cho}},
  \bibinfo {author} {\bibfnamefont {D.}~\bibnamefont {Chatzopoulos}}, \bibinfo
  {author} {\bibfnamefont {M.}~\bibnamefont {Leeuwenhoek}}, \bibinfo {author}
  {\bibfnamefont {C.}~\bibnamefont {Koks}},\ and\ \bibinfo {author}
  {\bibfnamefont {M.~P.}\ \bibnamefont {Allan}},\ }\bibfield  {title} {\bibinfo
  {title} {Imaging doubled shot noise in a josephson scanning tunneling
  microscope},\ }\href {https://doi.org/10.1103/PhysRevB.100.104506} {\bibfield
   {journal} {\bibinfo  {journal} {Phys. Rev. B}\ }\textbf {\bibinfo {volume}
  {100}},\ \bibinfo {pages} {104506} (\bibinfo {year} {2019})}\BibitemShut
  {NoStop}%
\bibitem [{\citenamefont {Yang}\ \emph {et~al.}(2013)\citenamefont {Yang},
  \citenamefont {Hebestreit}, \citenamefont {Josberger},\ and\ \citenamefont
  {Raschke}}]{Yang2013}%
  \BibitemOpen
  \bibfield  {author} {\bibinfo {author} {\bibfnamefont {H.~U.}\ \bibnamefont
  {Yang}}, \bibinfo {author} {\bibfnamefont {E.}~\bibnamefont {Hebestreit}},
  \bibinfo {author} {\bibfnamefont {E.~E.}\ \bibnamefont {Josberger}},\ and\
  \bibinfo {author} {\bibfnamefont {M.~B.}\ \bibnamefont {Raschke}},\
  }\bibfield  {title} {\bibinfo {title} {{A cryogenic scattering-type scanning
  near-field optical microscope}},\ }\href {https://doi.org/10.1063/1.4789428}
  {\bibfield  {journal} {\bibinfo  {journal} {Review of Scientific
  Instruments}\ }\textbf {\bibinfo {volume} {84}},\ \bibinfo {pages} {023701}
  (\bibinfo {year} {2013})}\BibitemShut {NoStop}%
\bibitem [{\citenamefont {McLeod}\ \emph {et~al.}(2017)\citenamefont {McLeod},
  \citenamefont {van Heumen}, \citenamefont {Ramirez}, \citenamefont {Wang},
  \citenamefont {Saerbeck}, \citenamefont {Guenon}, \citenamefont {Goldflam},
  \citenamefont {Anderegg}, \citenamefont {Kelly}, \citenamefont {Mueller},
  \citenamefont {Liu}, \citenamefont {Schuller},\ and\ \citenamefont
  {Basov}}]{Mcleod2017}%
  \BibitemOpen
  \bibfield  {author} {\bibinfo {author} {\bibfnamefont {A.~S.}\ \bibnamefont
  {McLeod}}, \bibinfo {author} {\bibfnamefont {E.}~\bibnamefont {van Heumen}},
  \bibinfo {author} {\bibfnamefont {J.~G.}\ \bibnamefont {Ramirez}}, \bibinfo
  {author} {\bibfnamefont {S.}~\bibnamefont {Wang}}, \bibinfo {author}
  {\bibfnamefont {T.}~\bibnamefont {Saerbeck}}, \bibinfo {author}
  {\bibfnamefont {S.}~\bibnamefont {Guenon}}, \bibinfo {author} {\bibfnamefont
  {M.}~\bibnamefont {Goldflam}}, \bibinfo {author} {\bibfnamefont
  {L.}~\bibnamefont {Anderegg}}, \bibinfo {author} {\bibfnamefont
  {P.}~\bibnamefont {Kelly}}, \bibinfo {author} {\bibfnamefont
  {A.}~\bibnamefont {Mueller}}, \bibinfo {author} {\bibfnamefont {M.~K.}\
  \bibnamefont {Liu}}, \bibinfo {author} {\bibfnamefont {I.~K.}\ \bibnamefont
  {Schuller}},\ and\ \bibinfo {author} {\bibfnamefont {D.~N.}\ \bibnamefont
  {Basov}},\ }\bibfield  {title} {\bibinfo {title} {Nanotextured phase
  coexistence in the correlated insulator {$V_2O_3$}},\ }\href
  {https://doi.org/10.1038/nphys3882} {\bibfield  {journal} {\bibinfo
  {journal} {Nature Physics}\ }\textbf {\bibinfo {volume} {13}},\ \bibinfo
  {pages} {80} (\bibinfo {year} {2017})}\BibitemShut {NoStop}%
\end{thebibliography}%

\newpage

\makeatletter

\setcounter{equation}{0}
\setcounter{figure}{0}
\setcounter{table}{0}
\setcounter{page}{1}
\setcounter{section}{0}
\renewcommand{\thesection}{S\Roman{section}}
\renewcommand{\theequation}{S\arabic{equation}}
\renewcommand{\thefigure}{S\arabic{figure}}
\renewcommand{\thetable}{S\Roman{table}}

\makeatother

\title{Supplementary Material: Local condensation of charge-$4e$ superconductivity
		at a nematic domain wall}
\author{Matthias Hecker }
\affiliation{School of Physics and Astronomy, University of Minnesota, Minneapolis
		55455 MN, USA}
\author{Rafael M. Fernandes}
\affiliation{School of Physics and Astronomy, University of Minnesota, Minneapolis
		55455 MN, USA}
\date{\today }
\maketitle

\onecolumngrid
\begin{center}
	\textbf{\large Supplementary Material: Local condensation of charge-$4e$ superconductivity
		at a nematic domain wall} 
\end{center}
\vspace{1cm}
\twocolumngrid

\section{Derivation of the variational free energy}
	
	Here, we derive the variational free energy associated with the ansatz
	$\mathcal{S}_{0}$ of the main text, i.e. we evaluate equation (3)
	of the main text. For convenience, we first repeat the setup of the
	problem and the notations introduced in the main text. Expressing
	the two-component superconducting order parameter $\boldsymbol{\Delta}=(\Delta_{1},\Delta_{2})$
	in terms of the four-component Nambu basis $\hat{\boldsymbol{\Delta}}=\left(\boldsymbol{\Delta},\bar{\boldsymbol{\Delta}}\right)$,
	we can rewrite the real-valued and complex-valued bilinear combinations
	as 
	\begin{align}
	\Psi^{n} & =\hat{\boldsymbol{\Delta}}^{\dagger}M^{n}\hat{\boldsymbol{\Delta}}, & \psi^{n} & =\hat{\boldsymbol{\Delta}}^{\dagger}m^{n}\hat{\boldsymbol{\Delta}}.\label{eq:Psipsi_SM}
	\end{align}
	Here, we defined the matrices 
	\begin{align}
	M^{A_{1g}} & =\tau^{0}\sigma^{0}/2, & m^{A_{1g}} & =\tau^{0}\sigma^{-}, & M^{A_{2g}} & =\tau^{y}\sigma^{z}/2,\nonumber \\
	M^{B_{1g}} & =\tau^{z}\sigma^{0}/2, & m^{B_{1g}} & =\tau^{z}\sigma^{-},\nonumber \\
	M^{B_{2g}} & =\tau^{x}\sigma^{0}/2, & m^{B_{2g}} & =\tau^{x}\sigma^{-},\label{eq:M_m-1}
	\end{align}
	with $\sigma^{\pm}=(\sigma^{x}\pm\mathsf{i}\sigma^{y})/2$ and $\tau^{i}$,
	$\sigma^{i}$ acting respectively on the internal superconducting
	subspace and on the Nambu space. For our specific setting of a one-dimensional
	grid of length $L$ with lattice sites labeled $i,j=1,\dots,N$, the
	superconducting action becomes 
	\begin{align*}
	\mathcal{S} & =\frac{L}{T}\sum_{i,j}\boldsymbol{\Delta}_{i}^{\dagger}\left[r_{0}\delta_{ij}+\frac{1}{2}f_{ij}^{0}\right]\tau^{0}\boldsymbol{\Delta}_{j}\;+\;\mathcal{S}^{\mathrm{int}}\,,
	\end{align*}
	where we introduced the gradient term
	\begin{align}
	\mathcal{S}^{\mathrm{grad}} & =\frac{L}{2T}\sum_{i,j}\boldsymbol{\Delta}_{i}^{\dagger}\tau^{0}f_{ij}^{0}\boldsymbol{\Delta}_{j},\label{eq:S_grad_SM}
	\end{align}
	described in terms of the hopping function $f_{ij}^{0}=\frac{t_{0}}{2}\left(2\delta_{ij}-\delta_{i,j+1}-\delta_{i,j-1}\right)$
	and the stiffness parameter $t_{0}>0$. Recall that $r_{0}=a_{0}(T-T_{0})$
	denotes the bare superconducting transition temperature with $a_{0},T_{0}>0$.
	The interaction part is given by \citep{Hecker2023} (see also Eq.
	(4) in main text)
	\begin{align}
	\mathcal{S}^{\mathrm{int}} & =\frac{L}{T}\sum_{i}\Big[u\,(\Psi_{i}^{A_{1g}})^{2}+v\,(\Psi_{i}^{A_{2g}})^{2}+w\,(\Psi_{i}^{B_{1g}})^{2}\Big],\label{eq:S_int_SM}
	\end{align}
	where the Landau parameters satisfy the conditions $u>0$ and $v,w>-u$,
	in order for the action to be bounded. 
	
	As discussed in the main text, within the Gaussian variational approach,
	we choose a trial action 
	\begin{align}
	\mathcal{S}_{0} & =\frac{1}{2}\frac{L}{T}\sum_{i,j}\hat{\boldsymbol{\Delta}}_{i}^{\dagger}\,\mathcal{G}_{i,j}^{-1}\,\hat{\boldsymbol{\Delta}}_{j},\label{eq:S0_SM}
	\end{align}
	that is characterized by the inverse Green's function 
	\begin{align}
	\mathcal{G}_{ij}^{-1} & =G_{i}^{-1}\,\delta_{ij}+f_{ij}^{0}M^{A_{1g}},\label{eq:var_Greens_fcn}\\
	G_{i}^{-1} & =2R_{i}M^{A_{1g}}+2\sum_{n\in\mathbb{G}_{\mathbb{R}}}\Phi_{i}^{n}M^{n}+\sum_{n\in\mathbb{G}_{\mathbb{C}}}\left(\bar{\phi}_{i}^{n}m^{n}+\mathrm{h.c.}\right),\label{eq:var_Greens_fcn2_SM}
	\end{align}
	which contains all variational parameters. Recall that we use $\Phi_{i}^{n}$
	for real-valued variational composite order parameters and $(\phi_{i}^{n},\bar{\phi}_{i}^{n})$
	for the complex-valued ones, where $n$ denotes the irreducible representation
	(IR) according to which the composite transforms. Moreover, we also
	define the mass renormalization parameter $R_{i}=r_{0}+\Phi_{i}^{A_{1g}}$.
	For convenience of notation, we introduce the IR sets $\mathbb{G}_{\mathbb{R}}=\{A_{2g},B_{1g},B_{2g}\}$
	and $\mathbb{G}_{\mathbb{C}}=\{A_{1g},B_{1g},B_{2g}\}$ in Eq. (\ref{eq:var_Greens_fcn2_SM}).
	Note that the local inverse Green's function (\ref{eq:var_Greens_fcn2_SM})
	is identical to Eq. (5) in the main text. 
	
	Since the trial action (\ref{eq:S0_SM}) is Gaussian, it is straightforward
	to evaluate the variational free energy {[}Eq. (6) in the main text{]}
	\begin{align}
	F_{v} & =-T\log Z_{0}+T\langle\mathcal{S}-\mathcal{S}_{0}\rangle_{0},\label{eq:Fv_SM}
	\end{align}
	with $Z_{0}=\int D(\boldsymbol{\Delta},\bar{\boldsymbol{\Delta}})\,e^{-\mathcal{S}_{0}}$
	being the partition function of the trial action, see for example
	Ref. \citep{Hecker2023} for technical details. In the real-space
	representation (\ref{eq:S0_SM}), it is convenient to promote the
	$4$-component field $\hat{\boldsymbol{\Delta}}_{i}$ to a $(4N)$-component
	field via $\doublehat{\boldsymbol{\Delta}}=\sum_{i}\hat{P}_{i}\hat{\boldsymbol{\Delta}}_{i}$,
	where the projector $\hat{P}_{i}$ is a $(4N)\times4$ dimensional
	matrix whose elements are either $0$ or $1$, and $\hat{P}_{i}^{T}\hat{P}_{j}=\delta_{ij}\mathbbm{1}_{4}$.
	The resulting variational free energy $F_{v}$, up to an unimportant
	constant, is given by
	\begin{align}
	F_{v} & =\frac{T}{2}\log\det\Big(\doublehat{\mathcal{G}}^{-1}\Big)+T\!\sum_{i}\!\Big\{2\big[r_{0}-R_{i}\!+\tilde{U}_{A_{1g}}G_{ii}^{A_{1g}}\big]G_{ii}^{A_{1g}}\nonumber \\
	& -2\sum_{n\in\mathbb{G}_{\mathbb{R}}}\left(\Phi_{i}^{n}\!-\tilde{U}_{n}G_{ii}^{n}\right)G_{ii}^{n}-\sum_{n\in\mathbb{G}_{\mathbb{C}}}\big[\left(\phi_{i}^{n}-\tilde{u}_{n}g_{ii}^{n}\right)\bar{g}_{ii}^{n}+\mathrm{c.c.}\big]\!\Big\}.\label{eq:var_free_en}
	\end{align}
	Here, $\doublehat{\mathcal{G}}$ is the inverse of $\doublehat{\mathcal{G}}^{-1}=\sum_{i,j}\hat{P}_{i}\mathcal{G}_{ij}^{-1}\hat{P}_{j}^{T}$
	and $G_{ij}^{n}$ is given by the decomposition of $\mathcal{G}_{ij}=\hat{P}_{i}^{T}\doublehat{\mathcal{G}}\hat{P}_{j}$
	onto the symmetry channels according to
	\begin{align}
	\mathcal{G}_{ij} & =2\!\!\!\sum_{n\in\{A_{1g},\mathbb{G}_{\mathbb{R}}\}}\!\!G_{ij}^{n}M^{n}+\!\!\sum_{n\in\mathbb{G}_{\mathbb{C}}}\!\!\Big[g_{ij}^{n}(m^{n})^{\dagger}\!+\bar{g}_{ij}^{n}m^{n}\Big].\label{eq:Greens_mat_projected}
	\end{align}
	Finally, in Eq. (\ref{eq:var_free_en}), the effective interaction
	parameters $\left\{ \tilde{U}_{n},\tilde{u}_{n}\right\} =\frac{T}{L}\left\{ U_{n},u_{n}\right\} $
	in each symmetry channel are given by
	\begin{align}
	U_{A_{1g}} & =3u+v+w, & u_{A_{1g}} & =u-v+w, & U_{A_{2g}} & =u+3v-w,\nonumber \\
	U_{B_{1g}} & =u-v+3w, & u_{B_{1g}} & =u+v+w,\nonumber \\
	U_{B_{2g}} & =u-v-w, & u_{B_{2g}} & =u+v-w,\label{eq:int_parameters}
	\end{align}
	which agree with the expressions found in the bulk case \citep{Hecker2023}. 
	
	\section{details of the free energy minimization}
	
	In this section, we present a few technical details related to the
	minimization of the free energy (\ref{eq:var_free_en}). As emphasized
	in the main text, we model the domain wall through the boundary conditions
	$\Phi_{1}^{B_{1g}}=-\Phi_{N}^{B_{1g}}=\Phi_{0}^{B_{1g}}$, $R_{1}=R_{N}=R_{0}$,
	and all other fields $\Phi_{1,N}^{n}=\phi_{1,N}^{n}=0$, where $\Phi_{0}^{B_{1g}}$
	and $R_{0}$ are the corresponding bulk values. 
	
	The free energy (\ref{eq:var_free_en}) effectively depends only on
	the parameters $T$, $\hat{t}_{0}\equiv t_{0}/\sqrt{uT/L}$, $\{v,w\}/u$,
	and $\{r_{0},R_{i},\Phi_{i}^{n},\phi_{i}^{n}\}/t_{0}$. In the spirit
	of the Ginzburg-Landau approach, we assume the most important temperature
	dependence to be in $r_{0}=a_{0}(T-T_{0})$, and we set $T=T_{0}$
	elsewhere, with $T_{0}$ denoting the bare SC transition temperature.
	To facilitate the minimization of the free energy (\ref{eq:var_free_en}),
	we supply the minimizer with the gradient expressions given by
	\begin{align}
	\frac{\partial F_{v}}{\partial X_{i}} & =T\!\sum_{j}\!\Bigg\{ V_{j}^{A_{1g}}\frac{\partial G_{jj}^{A_{1g}}}{\partial X_{i}}+\sum_{n\in\mathbb{G}_{\mathbb{R}}}V_{j}^{n}\frac{\partial G_{jj}^{n}}{\partial X_{i}}\nonumber \\
	& \quad+\sum_{n\in\mathbb{G}_{\mathbb{C}}}\left(v_{j}^{n}\frac{\partial g_{jj}^{n}}{\partial X_{i}}+\mathrm{c.c.}\right)\Bigg\},\label{eq:dFdX_SM}
	\end{align}
	where $X_{i}\in\{R_{i},\Phi_{i}^{n},\phi_{i}^{n},\bar{\phi}_{i}^{n}\}$
	can be any of the variational parameters. In writing Eq. (\ref{eq:dFdX_SM}),
	we exploited the fact that the partial derivatives $\frac{\partial F_{v}}{\partial X_{i}}\big|_{G^{n},g^{n}}=0$
	vanish \citep{Hecker2023}. Moreover, we defined 
	\begin{align}
	V_{j}^{A_{1g}} & =2\big[r_{0}-R_{j}\!+2\tilde{U}_{A_{1g}}G_{jj}^{A_{1g}}\big], & v_{j}^{n} & =-\left(\bar{\phi}_{j}^{n}-2\tilde{u}_{n}\bar{g}_{jj}^{n}\right),\nonumber \\
	V_{j}^{n} & =-2\left(\Phi_{j}^{n}\!-2\tilde{U}_{n}G_{jj}^{n}\right).\label{eq:Vn}
	\end{align}
	
	To determine the remaining derivatives in Eq. (\ref{eq:dFdX_SM}),
	we use the Green's function (\ref{eq:Greens_mat_projected}) to identify
	\begin{align}
	G_{jj}^{n} & =\frac{1}{2}\mathrm{tr}\left[\mathcal{G}_{jj}M^{n}\right], & g_{jj}^{n} & =\frac{1}{2}\mathrm{tr}\left[\mathcal{G}_{jj}m^{n}\right].\label{eq:Gngn}
	\end{align}
	Then, using the introduced projector matrix $\hat{P}_{i}$, we obtain
	the relationship
	\begin{align}
	\frac{\partial\mathcal{G}_{jj}}{\partial X_{i}} & =-\hat{P}_{j}^{T}\doublehat{\mathcal{G}}\frac{\partial\doublehat{\mathcal{G}}^{-1}}{\partial X_{i}}\doublehat{\mathcal{G}}\hat{P}_{j}=-\sum_{i_{1}i_{2}}\mathcal{G}_{ji_{1}}\frac{\partial\mathcal{G}_{i_{1}i_{2}}^{-1}}{\partial X_{i}}\mathcal{G}_{i_{2}j},\label{eq:dGdX}
	\end{align}
	which leads to the expressions
	\begin{align}
	\frac{\partial\mathcal{G}_{jj}}{\partial R_{i}} & =-2\mathcal{G}_{ji}M^{A_{1g}}\mathcal{G}_{ij}, & \frac{\partial\mathcal{G}_{jj}}{\partial\Phi_{i}^{n}} & =-2\mathcal{G}_{ji}M^{n}\mathcal{G}_{ij},\nonumber \\
	\frac{\partial\mathcal{G}_{jj}}{\partial\bar{\phi}_{i}^{n}} & =-\mathcal{G}_{ji}m^{n}\mathcal{G}_{ij}.\label{eq:dGdR_SM}
	\end{align}
	It is now straightforward to compute the derivatives in Eq. (\ref{eq:Gngn})
	by using Eq. (\ref{eq:dGdR_SM}); we find, for instance
	\begin{align}
	\frac{\partial G_{jj}^{n}}{\partial R_{i}} & =-\mathrm{tr}\left[\mathcal{G}_{ji}M^{A_{1g}}\mathcal{G}_{ij}M^{n}\right],\nonumber \\
	\frac{\partial g_{jj}^{n}}{\partial R_{i}} & =-\mathrm{tr}\left[\mathcal{G}_{ji}M^{A_{1g}}\mathcal{G}_{ij}m^{n}\right],\label{eq:dGdR2_SM}
	\end{align}
	and similar expressions for the other variational parameters.

\end{document}